\documentclass[8.5pt,twoside,twocolumn]{article}
\usepackage[super,sort&compress,comma]{natbib} 
\usepackage{mhchem}
\usepackage{times,mathptmx}
\usepackage{sectsty}
\usepackage{balance} 

\usepackage{graphicx} 
\usepackage{lastpage}
\usepackage{array}
\usepackage{color}
\usepackage[format=plain,justification=raggedright,singlelinecheck=false,font=small,labelfont=bf,labelsep=space]{caption} 
\usepackage{fancyhdr}
\newcommand{\am}{\mbox{-3.10$\pm0.06$} }

\begin{document}

\thispagestyle{plain}
\fancypagestyle{plain}{
\renewcommand{\headrulewidth}{1pt}}
\renewcommand{\thefootnote}{\fnsymbol{footnote}}
\renewcommand\footnoterule{\vspace*{1pt}%
\hrule width 3.4in height 0.4pt \vspace*{5pt}} 
\setcounter{secnumdepth}{5}

\makeatletter 
\def\subsubsection{\@startsection{subsubsection}{3}{10pt}{-1.25ex plus -1ex minus -.1ex}{0ex plus 0ex}{\normalsize\bf}} 
\def\paragraph{\@startsection{paragraph}{4}{10pt}{-1.25ex plus -1ex minus -.1ex}{0ex plus 0ex}{\normalsize\textit}} 
\renewcommand\@biblabel[1]{#1}            
\renewcommand\@makefntext[1]%
{\noindent\makebox[0pt][r]{\@thefnmark\,}#1}
\makeatother 
\renewcommand{\figurename}{\small{Fig.}~}
\sectionfont{\large}
\subsectionfont{\normalsize} 

\fancyfoot{}
\fancyfoot[RO]{\footnotesize{\sffamily{1--\pageref{LastPage} ~\textbar  \hspace{2pt}\thepage}}}
\fancyfoot[LE]{\footnotesize{\sffamily{\thepage~\textbar\hspace{3.45cm} 1--\pageref{LastPage}}}}
\fancyhead{}
\renewcommand{\headrulewidth}{1pt} 
\renewcommand{\footrulewidth}{1pt}
\setlength{\arrayrulewidth}{1pt}
\setlength{\columnsep}{6.5mm}
\setlength\bibsep{1pt}

\twocolumn[
  \begin{@twocolumnfalse}
\noindent\LARGE{\textbf{Quantum Monte Carlo for Noncovalent Interactions: An Efficient Protocol Attaining Benchmark Accuracy}}
\vspace{0.6cm}

\noindent\large{\textbf{Mat\'u\v{s} Dubeck\'{y},\textit{$^{a,}$}$^{\ast}$
Ren\'e Derian,\textit{$^{b}$}
Petr Jure\v{c}ka,\textit{$^{a,}$}$^{\ast}$
Lubos Mitas,\textit{$^{c}$}
Pavel Hobza,\textit{$^{d,a}$}
Michal Otyepka\textit{$^{a}$}}
}\vspace{0.5cm}



\noindent \normalsize{Reliable theoretical predictions of noncovalent interaction energies, which are important e.g. in drug-design and hydrogen-storage applications, belong to longstanding challenges of contemporary quantum chemistry. In this respect, the fixed-node diffusion Monte Carlo (FN-DMC) is a promising alternative to the commonly used ``gold standard'' coupled-cluster CCSD(T)/CBS method for its benchmark accuracy and favourable scaling, in contrast to other correlated wave function approaches.
This work is focused on the analysis of protocols and possible tradeoffs for FN-DMC estimations of noncovalent interaction energies and
proposes an efficient yet accurate computational protocol using
simplified explicit correlation terms with a favorable $O(N^3)$ scaling.
It achieves an excellent agreement (mean unsigned error $\sim$0.2 kcal/mol) with respect to the CCSD(T)/CBS data on a number of complexes, including benzene/hydrogen, \mbox{T-shape} benzene dimer, stacked adenine-thymine and a set of small noncovalent complexes A24. 
The high accuracy and reduced computational costs predestinate the reported protocol for practical interaction energy calculations of large noncovalent complexes, where the CCSD(T)/CBS is prohibitively expensive.
}
\vspace{0.5cm}
 \end{@twocolumnfalse}
  ]

\section{Introduction}
\footnotetext{\textit{$^{a}$}~Regional Centre of Advanced Technologies and Materials,
Department of Physical Chemistry, Faculty of Science, Palack\'y University
Olomouc, t\v{r}.~17~listopadu 12, 771 46 Olomouc, Czech Republic.}
\footnotetext{\textit{$^{b}$}~Institute of Physics, Slovak Academy of Sciences, D\'ubravsk\'a cesta 9, 845~11 Bratislava, Slovakia.}
\footnotetext{\textit{$^{c}$}~Department of Physics and CHiPS, North Carolina State University, Raleigh, NC 27695, USA.}
\footnotetext{\textit{$^{d}$}~Institute of Organic Chemistry and Biochemistry, Academy of Sciences of the Czech Republic, Flemingovo n\'am. 2, 166 10 Prague 6, Czech Republic.}
\footnotetext{\textit{$^{*}$} E-mail: matus.dubecky@upol.cz, petr.jurecka@upol.cz}

Noncovalent interactions between molecular complexes and/or their parts are of key importance in many areas of chemistry, biology and materials science~\cite{Hobzabk,Riley2010rev,Grimme2006}.
Their striking manifestations include the properties of liquids, molecular recognition~\cite{Salonen2011} or the structure and function of bio-macromolecules, to name just a few~\cite{Georgakilas2012rev,Riley2010rev}. 
In general, experiments make it possible to obtain information on the strength of noncovalent interactions e.g. from dissociation and adsorption enthalpies. Nevertheless, direct information on their nature is usually unavailable. In order to characterise the noncovalent interaction/s of interest more precisely, one typically resorts to a combination of multiple techniques~\cite{Lazar2013,Hobzabk}. 
Theory and high-accuracy calculations here usefully complement experiments by providing the detailed information necessary for their fundamental understanding. 
As the binding energies of noncovalent complexes are weak (typically 0.5 to 30 kcal/mol in small complexes) when compared to a typical covalent bond energy (100~kcal/mol), their computations require methods of exceptional quality. For benchmark purposes, only approaches with a degree of accuracy beyond the accepted chemical accuracy (1~kcal/mol) are considered reliable enough. Ideally, the level of subchemical accuracy~\cite{Schuurman2004} (0.1 kcal/mol) should be reached. This task poses a long-standing challenge to modern computational chemistry.

The established ``gold standard'' of the quantum mechanical calculations of noncovalent interaction energies~\cite{Hobza2012rev} has been the CCSD(T)~\cite{Raghavachari1989} (coupled-cluster singles and doubles with perturbative triples) method, which guarantees the desired accuracy, provided that large enough basis sets and/or complete basis set (CBS) 
extrapolations are employed. The benchmark capability of CCSD(T)/CBS has recently been verified on a set of small noncovalent complexes by a higher-order CCSDT(Q) method~\cite{Rezac2013} and in small bases also by the \mbox{CCSDTQ}, \mbox{CCSDTQP} and 
full CI (configuration interaction)~\cite{Simova2013}. Nevertheless, because of the rapid growth 
of the CCSD(T) computational cost in the basis set size $M$, $\propto O(M^7)$, and even the more steeply growing demands 
of higher-order coupled-cluster approaches, their practical use remains limited to relatively small systems~\cite{Jurecka2006,Podeszwa2010,Takatani2010,Rezac2011,Brndiar2012}. For summary of developments in this field, cf. Ref.~\citenum{Gordon2012rev}.  

A promising alternative to solve the non-relativistic
Schr\"{o}dinger equation for electrons in Born-Oppenheimer  approximation is the fixed-node 
(FN) diffusion Monte Carlo (DMC/FN-DMC) method, a member of the quantum
Monte Carlo (QMC) class of methods based on random sampling. FN-DMC solves an imaginary-time Schr\"{o}dinger equation projecting out the exact ground-state within the constraints given by the nodal surface ($\Psi_T=0$) of the best available trial wave function ($\Psi_T$), required to preserve the antisymmetry of the simulated electronic state~\cite{Foulkes2001rev,Austin2012rev}. The FN-DMC approach is favourable 
for its direct treatment of many-body correlations (competitive to high-order approaches including full CI)~\cite{Foulkes2001rev,Austin2012rev}, 
low-order polynomial scaling \mbox{$\propto O(N^{3-4})$}~\cite{Foulkes2001rev} with the number of electrons $N$ and its intrinsic
massive parallelism~\cite{Foulkes2001rev,Austin2012rev}. 
The representation of a
wave function as an ensemble of walkers (electron real space position vectors) in QMC 
results in a CBS-equivalent mode  and energies that are insensitive to basis set superposition errors (possibly present in $\Psi_T$)~\cite{Austin2012rev}. QMC makes it possible to sample sophisticated many-body wave function ans\"atze with explicit inter-electronic
dependencies efficiently, as the wave functions are sampled stochastically~\cite{Jastrow1955,Foulkes2001rev,Bajdich2009rev,Austin2012rev}.
On the other hand, it suffers from disadvantages like slowly convergent error bars intrinsic to stochastic sampling, the need to circumvent
the difficulties in sampling antisymmetric wave functions and/or complicated estimations of quantities beyond energies. Despite the mentioned limitations, QMC approaches are behind some of 
the most paradigmatic results obtained for quantum systems such as the correlation energy
of electron gas~\cite{Ceperley1980}, widely used in computational electronic structure methods. For more details
and background on QMC, we refer the reader to the published reviews~\cite{Foulkes2001rev,Austin2012rev,Bajdich2009rev}.

In the domain of noncovalent interactions, there are two salient features of FN-DMC. First, 
the method accurately describes the dynamic correlation effects crucial for  
noncovalent interactions, since it recovers
{\it all possible many-body correlations} within the constraints given by the nodal surface~\cite{Foulkes2001rev}. Second, the finite
FN error of $\Psi_T$ used in DMC is expected to cancel out in energy differences nearly exactly~\cite{Mella2003,Diedrich2005,Korth2008,Dubecky2013}, because the nodes in the region of the molecule $A$ are essentially unchanged by 
the presence of the weakly interacting molecule $B$ 
and vice-versa. This trend has been qualitatively confirmed by a direct inspection of nodal surfaces~\cite{Dubecky2013}.
For these reasons, FN-DMC based on single-determinant trial functions (containing e.g. Kohn-Sham orbitals) has been found very promising in small/medium~\cite{Benedek2006,Korth2008,Dubecky2013} and medium/large molecular systems~\cite{Grimme2006,Santra2008,Ma2009,Tkatchenko2012,Hongo2010,Hongo2013,Alfe2013,Ambrosetti2014}, where conventional correlated methods are inapplicable due to the
prohibitive computational cost.
 { Note that the requirement of small error bars, a must in noncovalent interactions, leads to a large cost (large prefactor) of the calculations. For example, calculations of small complexes like those considered as a teaching set in the current work are much more costly in QMC than in CCSD(T)/CBS. Nevertheless, the situation turns to the contrary in large systems.}
Although systematic understanding of error cancellation in QMC energy differences is still far from complete,
the data accumulated to date begin to delineate the method performance for various types of systems \cite{Petruzielo2012,Dubecky2013,Rasch2013}. 
For more detailed presentations, see Refs.~\citenum{Austin2012rev,Mella2003,Korth2008} and \citenum{Dubecky2013}.

In our previous study in this field~\cite{Dubecky2013}, we demonstrated the ability of the \mbox{FN-DMC} to reach the CCSD(T)/CBS within \mbox{0.1~kcal/mol} on a teaching set of six small noncovalent complexes: the dimers of ammonia, water, hydrogen fluoride, methane, ethene and ethene/ethyne complex.
The successive procedure was subsequently tested on complexes of benzene/methane, benzene/water and  \mbox{T-shape} benzene dimer, where the FN-DMC deviated by no more than 0.25~kcal/mol with respect to the best available CCSD(T)/CBS interaction energy estimates. In a way, QMC performed much better than expected a priori~\cite{Korth2008}. 
Note that the genuine accuracy of CCSD(T) for these sizable systems has not yet been confirmed to subchemical accuracy at a higher level (CCSDT(Q)), and both approaches therefore provide independent interaction energy estimates that agree remarkably well. In addition, extended basis sets (e.g. aug-cc-pVTZ) are unavoidable in these CC calculations, because small bases lead to qualitatively incorrect results~\cite{Rezac2013}. For more of the complexes considered in the current work, such CC calculations are prohibitively expensive and if the FN-DMC calculations prove to be accurate enough, this route immediately provides new means for adressing of such difficult problems. 

In the present work, we analyze the possible FN-DMC approaches and tradeoffs and report on a considerably cheaper and yet accurate scheme based on a two-body explicit correlation (Jastrow) factor, which is a simplification with respect to the traditionally used three-body Jastrow term. The discovery of its ability to attain $~\sim$0.2~kcal/mol accuracy in case of noncovalent interactions as reported in the current work represents an important advance. The favourable performance of the simplified scheme also directly confirms the previous observation that the fixed-node error cancellation is the main reason for the success of the one-determinant FN-DMC in weakly bound noncovalent complexes~\cite{Dubecky2013}.

\section{Molecular Complexes}
{The calculations reported below were performed on a diverse set of hydrogen-bonded and/or dispersion-bound complexes for which reliable estimates of interaction energies have been published~\cite{Jurecka2006,Takatani2010,Rezac2013}. The teaching set consists of the dimers of ammonia, water, hydrogen fluoride, methane, ethene and the ethene/ethyne complex as in our previous work to allow a convenient comparison. The complexes used for testing purposes include  benzene/H$_2$, benzene/methane, benzene/water, \mbox{T-shape} benzene dimer, stacked adenine-thymine complex and the whole set of 24 small noncovalent complexes A24~\cite{Rezac2013}. 

The selection of the benzene/H$_2$ is motivated by our interest in assessment of interaction energies between carbon-based materials and H$_2$, useful for hydrogen-storage applications. The adenine-thymine complex, on the other hand, serves as a first stringent test of our approach in DNA base-pair interactions, and for future reference in the assessment of interactions in larger DNA fragments.}

\section{Analysis of QMC Protocols}
\begin{table*}[ht!]
\scriptsize
\caption{The FN-DMC interaction energies $E$ (kcal/mol) obtained from various tested protocols compared to the CCSD(T)/CBS reference $E_{R}$ (kcal/mol). The protocol-type attribute, if applicable, is indicated in the column P. For clarity, analyzed features of the protocols with respect to the standard~\cite{Dubecky2013} 3tJ protocol are indicated by the bold typeface. Abbreviations: $\Psi_T$ - trial wave function, LC - a linear combination of energy (95\%) and variance (5\%), $\Delta$ - the difference with respect to the reference, Dis. - a distinct Jastrow factor (see text).}
\centering
\begin{tabular}{l|c|c|ccc|c|ccc}
\hline
\hline
Complex       & $E_{R}$ & P  &  &$\Psi_T$&        & VMC opt. &  & FN-DMC &  \\
              &&    & Basis set & Method & Jastrow & cost function      & timestep/a.u. &  ECP treatment &  $E$    \\
\hline
Ammonia dimer & -3.15$^a$ &    & TZV       & B3LYP  & 3tJ                & LC    & 0.005 & T-moves & -3.33$\pm$0.07$^b$  \\
              &&    & QZV       & B3LYP  & 3tJ                & LC    & 0.005 & T-moves & -3.47$\pm$0.07$^b$  \\
              && 3tJ& aug-TZV   & B3LYP  & 3tJ                & LC    & 0.005 & T-moves & -3.10$\pm$0.06$^b$  \\
              &&    & \textbf{aug-QZV}
                                & B3LYP  &3tJ                 & LC    & 0.005 & T-moves & -3.13$\pm$0.07$^b$  \\
              &&    & aug-TZV   & \textbf{HF}&3tJ             & LC    & 0.005 & T-moves & -3.12$\pm$0.07$^b$  \\
              && 2tJ& aug-TZV   & B3LYP  & \textbf{2tJ}       & LC    & 0.005 & T-moves & -3.15$\pm$0.05 \\
              &&    & aug-TZV   & B3LYP  & \textbf{2tJ}       & LC    & \textbf{0.01} 
                                                                                           & T-moves & -3.14$\pm$0.05  \\
              &&    & aug-TZV   & B3LYP  & 3tJ                & \textbf{Variance}  & 0.005 & T-moves & -3.28$\pm$0.04   \\
              &&    & aug-TZV   & B3LYP  & 3tJ                & LC    & \textbf{0.01} 
                                                                                           & T-moves & -3.22$\pm$0.07  \\              
              &&    & aug-TZV   & B3LYP  & 3tJ                & LC    & 0.005 & \textbf{Locality} & -3.27$\pm$0.07   \\
              &&    & aug-TZV   & B3LYP  & 3tJ                & LC    & \textbf{0.01} & \textbf{Locality}
                                                                                                     & -3.39$\pm$0.06   \\
              \hline
Water  dimer  &-5.07$^a$& 3tJ& aug-TZV   & B3LYP  & 3tJ                & LC    & 0.005 & T-moves & -5.26$\pm$0.08$^b$   \\
              &&    & aug-TZV   & B3LYP  & 3tJ                & LC    & \textbf{0.01} 
                                                                                           & T-moves & -5.13$\pm$0.08      \\
              &&    & aug-TZV   & B3LYP  & \textbf{3tJ Dis.}  & LC    & 0.005 & T-moves & -5.15$\pm$0.08$^b$     \\
              &&    & aug-TZV   & B3LYP  & \textbf{3tJ Dis.}  & LC    & \textbf{0.01} & T-moves & -5.26$\pm$0.08   \\
              && 2tJ& aug-TZV   & B3LYP  & \textbf{2tJ}       & LC  & 0.005   & T-moves & -5.24$\pm$0.09  \\
              &&    & aug-TZV   & B3LYP  & \textbf{2tJ Dis.}  & LC  & 0.005   & T-moves & -5.50$\pm$0.08 \\
              &&    & aug-TZV   & B3LYP  & \textbf{2tJ}       & LC  & \textbf{0.01}  
  & T-moves & -5.33$\pm$0.09  \\
\hline
Methane dimer &-0.53$^a$& 3tJ& aug-TZV   & B3LYP  & 3tJ       & LC    & 0.005 & T-moves & -0.44$\pm$0.05$^b$   \\
              &&    & \textbf{aug-QZV}
                                & B3LYP  & 3tJ                & LC    & 0.005 & T-moves & -0.55$\pm$0.04  \\
              &&    & aug-TZV   & \textbf{HF}  &   3tJ        & LC    & 0.005 & T-moves & -0.52$\pm$0.10   \\
              && 2tJ& aug-TZV   & B3LYP  & \textbf{2tJ}       & LC    & 0.005 & T-moves & -0.60$\pm$0.07   \\
              &&    & aug-TZV   & \textbf{HF}  & \textbf{2tJ} & LC    & 0.005 & T-moves & -0.64$\pm$0.08   \\
\hline
Ethene dimer  &-1.48$^a$& 3tJ& aug-TZV   & B3LYP  & 3tJ       & LC    & 0.005 & T-moves & -1.47$\pm$0.09$^b$   \\
              &&    & \textbf{aug-QZV}
                                & B3LYP  & 3tJ                & LC    & 0.005 & T-moves & -1.54$\pm$0.09  \\
              &&    & aug-TZV   & B3LYP  & 3tJ                & LC  & \textbf{0.01}  
                                                                                           & T-moves & -1.42$\pm$0.09  \\
              && 2tJ& aug-TZV   & B3LYP  & \textbf{2tJ}       & LC  & 0.005   & T-moves & -1.54$\pm$0.09  \\
              &&    & aug-TZV   & B3LYP  & \textbf{2tJ}       & LC  & \textbf{0.01}
                                                                                           & T-moves & -1.55$\pm$0.09  \\
                                                     \hline
\hline
\end{tabular}
\begin{tabular}{l}
$^a$~Takatani et al.~\cite{Takatani2010},~ $^b$~Dubeck\'y et al.~\cite{Dubecky2013}
\end{tabular}
\label{tabrescmp}
\end{table*}
The development of a FN-DMC-based methodology leading to accurate noncovalent interaction energies involves extensive testing and elimination of biases that affect the final results. 
Indeed, this has to be done in a stepwise manner since several sets of parameters enter 
the multistage approach~\cite{Dubecky2011,Horvathova2012}: 

i) The ansatz for the trial wave function  $\Psi_T$ must be selected first.  The "standard model" and frequently used choice is the Slater-Jastrow~\cite{Foulkes2001rev,Austin2012rev} functional form, a product of determinant(s) and an explicit correlation term (Jastrow factor), as employed here. Then it is necessary to choose an effective core potential (ECP, if any) and a basis set (e.g., aug-TZV). Subsequently, the Slater determinant(s) are constructed with  DFT, Hartree-Fock~(HF) or post-HF orbitals. Finally, terms included in the Jastrow factor~\cite{Jastrow1955} must be specified. They include, for instance,  electron-electron (ee), electron-nucleus (eN), and electron-electron-nucleus (eeN) terms, which contain explicit functional dependence on inter-particle distances and thus efficiently describe dynamic correlation effects~\cite{Foulkes2001rev}.
 
ii) The variational (VMC) optimisation step consists of the selection of the VMC cost function and a parametric optimisation of $\Psi_T$, which may or may not include adjustment of the nodal surfaces (which we avoid in the curent work). It is possible to improve the nodes by the reoptimisation of the orbitals and/or determinant expansion coefficients.

iii) The final FN-DMC ground-state projection calculations depend, in addition to the $\Psi_T$ optimised in step ii), on the parameters of the DMC simulation itself, including an imaginary time step, the treatment of ECPs, the target walker population/s and target error bar/s, to name but a few most important.

In general, the parameters and/or choices in all points,  i) to iii), affect the final interaction 
energies obtained after the production DMC simulations in iii), as the differences of the statistically independent total-energy expectation values (with associated error bars). The parameters in i) and ii), in addition to the energies accumulated during DMC runs, modify also the energy variance, thus  determining the length of the DMC simulations to reach the fixed target statistical accuracy.

Tab.~\ref{tabrescmp} shows the representative setups and the related results that helped us to trace the importance of the parameter changes considered and identify useful protocols; here, the preceding one~\cite{Dubecky2013} is labelled 3tJ and the new one 2tJ. These differ in the number of terms (ee, eN and eeN, vs. ee and eN) considered in the Jastrow factor (cf. the Methods).  
The key observations from~Tab.~\ref{tabrescmp}  may be summarised as follows.

\subsection{Basis Sets}

In the ammonia dimer complex, where the reference interaction energy amounts to \mbox{-3.15~kcal/mol}~\cite{Takatani2010}, the TZV and QZV bases result in interaction energies of \mbox{-3.33$\pm0.07$} and \mbox{-3.47$\pm0.07$}~kcal/mol, whereas the aug-TZV and \mbox{aug-QZV} bases lead to FN-DMC interaction energies of \am and \mbox{-3.13$\pm0.07$~kcal/mol}~\cite{Dubecky2013}, respectively. The presence of augmentation functions in $\Psi_T$ is therefore crucial~\cite{Diedrich2005}, whereas an increase of the basis set cardinality beyond the TZV level seems to play a smaller role than in the standard methods of quantum chemistry. In the methane dimer and ethene dimer complexes, the \mbox{aug-TZV} data are found to be statistically indistinguishable from the \mbox{aug-QZV} data as well. Since the \mbox{aug-TZV} basis set reaches the reference data within 0.1~kcal/mol in the whole teaching set considered~\cite{Dubecky2013}, we have used it throughout the study. 

Note that the aug-QZV basis set with approximately two times more basis functions than the aug-TZV lowers the energy variance.  
For example, in the methane dimer complex, the variance is improved by 0.01 a.u. (from 0.19 to 0.18 a.u.), decreasing 
the sampling needed to reach the same error bar by about 10\%. Nevertheless, the cost of the aug-QZV calculations, for the fixed number of DMC steps is three times higher, making the \mbox{aug-TZV} approach still much more favourable when considering the overall cost/accuracy ratio.

\subsection{Orbitals in the Slater Part of $\Psi_T$}

Since the FN-DMC energies depend primarily on the nodal surface of the trial wave function $\Psi_T$, we tested HF and B3LYP sets of orbitals in Slater determinants.
 In the ammonia dimer complex~\cite{Dubecky2013}, we found that the HF and B3LYP orbitals provide FN-DMC interaction energies that are indistinguishable within the error bars, namely \mbox{-3.12$\pm0.07$} vs. {\mbox\am kcal/mol}, and both in good agreement with the CCSD(T)/CBS reference (\mbox{-3.15~kcal/mol}~\cite{Takatani2010}). Similar conclusions apply in the case of methane dimer, separately in both types of the schemes considered, 2tJ and 3tJ alike. The total energies from B3LYP orbitals were always found to produce variationally lower total energy expectation values\cite{Kolorenc2010,Per2012} than those from HF (e.g. in ammonia dimer by $\sim$0.001 a.u.). This indicates better quality of the B3LYP nodal surfaces and therefore we used B3LYP orbitals for the rest of the calculations as well. 
We note that the results are expected to depend on the choice of the orbitals only very weakly due 
to the favourable FN error cancellation that takes place in weakly interacting complexes~\cite{Mella2003,Diedrich2005,Korth2008,Dubecky2013}.

Since the concept of error cancellation is not limited to one determinant, similar behaviour is expected in case of {\it noncovalent interactions} between open-shell systems where complete-active-space wave functions capturing multi-reference effects~\cite{Schautz2004,Dubecky2010} may be used instead of Hartree-Fock/Kohn-Sham determinants.

\subsection{Jastrow Factor}

The considered variations of the Jastrow term include the reduction of terms (2tJ, cf. the Methods) with respect to the previously reported version (3tJ), and the so-called distinct Jastrow factor including distinct parameters (3tJ Dis.) on the non-equivalent atoms of the same type. We have found, that e.g. in ammonia, water and ethene, 2tJ and 3tJ protocols generate approximately the same results. On the other hand, this is not the case in the methane dimer. The water dimer is an example where the 3tJ Jastrow factor is not sufficient and the corresponding interaction energy (\mbox{-5.26$\pm$0.08~kcal/mol}) deviates considerably from the reference (\mbox{-5.07~kcal/mol~\cite{Takatani2010}}). In order to reach a subchemical accuracy margin, a distinct 3tJ Jastrow correlation factor must be considered (3tJ Dis.,\mbox{-5.15$\pm$0.08~kcal/mol}). On the other hand, the use of 2tJ scheme in combination with the distinct feature (2tJ Dis., Tab.~\ref{tabrescmp}) is (currently) not recommended, since in the studied 
case it produced the value that is too off (\mbox{-5.
50$\pm$0.08~kcal/mol}). The reason probably being an insufficient number of sampling points used in VMC optimisation~(cf. Ref.\citenum{Zen2013}). A reasonable compromise between the accuracy and cost is thus provided by the 2tJ scheme (\mbox{-5.24$\pm$0.09~kcal/mol}), which is acceptable within the less tight but acceptable target error criterion, e.g. \mbox{0.2-0.3~kcal/mol}.

\begin{figure}[t!]
 \centering
 \caption{The DMC timings to reach 0.1 kcal/mol error bar, in hours of running on 64 cores, versus the the total count of electrons $N$.~\newline}
 \includegraphics[width=210px]{timings.eps}
  \label{figtim}
\end{figure}

\begin{table}[t!]
\scriptsize
\caption{The timings (in hours of running on 64 cores) for VMC optimization $t_{\text{opt}}$, DMC (to reach 0.1 kcal/mol error bar) $t_{\text{DMC}}$ and total time $t$ to get a total single point energy. The $N$ denotes the total count of electrons.}
\centering
\begin{tabular}{lccccc}
\hline
\hline
Complex & $N$ & Protocol  & $t_{\text{opt}}$&$t_{\text{DMC}}$ &$t$\\
\hline
Water monomer & 8
 & 2tJ & 0.1 & 0.5 & 1.9 \\ 
&& 3tJ & 1.9 & 0.6 & 2.3 \\ 
\\
Water dimer & 16
 & 2tJ      & 0.2 &  10.5& 10.7\\ 
&& 3tJ      & 4.7 &  24.4& 29.1\\ 
&& 3tJ Dis. & 9.3 &  93.0& 102.3\\
\\

Benzene dimer & 60
 & 2tJ & 4& 616 & 620\\ 
&& 3tJ & 208 & 4931 & 5139\\ 
\hline\hline
\end{tabular}\\
\label{tabtim}
\end{table}

{The comparison of timings (on 64 cores) for calculations with various Jastrow factors, including VMC optimization, DMC to reach 0.1 kcal/mol error margin and total timings to get a single point total energy are reported in Tab.~\ref{tabtim}. In addition, the DMC timings are illustrated in  Fig.~\ref{figtim}. The data in Tab.~\ref{tabtim} show that the typical VMC optimization cost is negligible with respect to the DMC part of a typical calculation for noncovalent interaction energy purposes, i.e. the total timings are dominated by DMC. The DMC timings indicate that the 2tJ scheme cost scales as $N^3$ while the 3tJ approximately as $N^4$ (cf.~Fig.~\ref{figtim}), where $N$ is the number of electrons. Since the scaling of 2tJ approach is asymptotically much more favorable and generates results with comparable quality, we believe it will be very useful in calculations of large complexes where more complex Jastrow terms are very costly.}

\subsection{VMC Cost Function}

The variance used as a cost function in VMC optimization of the Jastrow term parameters leads to higher FN-DMC total 
energies when compared to energy minimisation (with 95\% of energy and 5\% of variance, cf. the Methods)~\cite{Umrigar2005}. For example, test calculations employing variance minimisation in ammonia lower the energy variance, as expected, but the total energy remains higher (by 0.0012 a.u. in dimer), and the interaction energy produced in this way (\mbox{-3.28$\pm0.1$~kcal/mol}) also deviates more from the reference. We thus recommend using a large fraction of energy in the VMC-optimisation cost function.
 
\subsection{DMC Time Step}

For completeness, we explore the DMC time step of 0.01 a.u. in addition to our standard conservative time-step setting of 0.005 a.u., used to avoid the extrapolation of energy to a zero time step~\cite{Korth2008}. A conclusive discussion of this point is, however, not possible because the error bars do not allow statements of statistical significance. Neither do we thus attempt zero time step extrapolations; we discuss the observations only qualitatively. In ammonia dimer, we have observed a deterioration of the final interaction energies in the case of 3tJ with an increase of the timestep, whereas in the case of 2tJ, no apparent dependence arises. In the case of water dimer, the calculation of energies using an increased time step shows that only the 3tJ Dis. scheme is able to approach the reference in the zero time step limit. In ethene dimer, there is no significant dependence on the time step in either of the cases considered (2tJ and 3tJ). The 
time step of 0.005 a.u. is accurate enough for our purposes and it is used throughout the study.
 
\subsection{ECP Treatment in DMC}

In order to reduce the numerical cost of the calculations, we have removed core electrons using ECPs. In ammonia dimer, we have found that the T-moves~\cite{Casula2006} scheme used to treat ECPs in DMC produces more accurate results than the locality approximation~\cite{Mitas1991} (where the error reaches \mbox{$\sim$0.2~kcal/mol}), as expected, and we have thus combined T-moves with a short time step (0.005 a.u.).

\subsection{Summary}

To summarise the discussion related to the analysis of protocols and tradeoffs relevant in calculations of noncovalent interactions (between closed-shell complexes), we conclude that for general setups, the main group elements and the target \mbox{$\sim$0.3~kcal/mol} accuracy (in small systems, or formally per one bond), the use of the following scheme (2tJ protocol) is recommended:
i) single-determinant trial wave functions of Slater-Jastrow type using B3LYP orbitals and an aug-TZV basis set,
ii) an exhaustively optimised Jastrow factor (keeping the Slater determinant intact) with ee, and eN terms, and, 
iii) a FN-DMC ground-state projection using the T-moves scheme~\cite{Casula2006} 
and a time step of 0.005~a.u. The error bars should be converged to \mbox{0.1-0.2~kcal/mol} to obtain statistically meaningful results. To this end, DMC projection times of several thousands of a.u. with large walker ensembles are unavoidable.

\begin{table*}[t!]
\scriptsize
\caption{The FN-DMC interaction energies $E$ (kcal/mol) obtained by the new protocol based on simplified Jastrow factor (2tJ, cf. section 3), compared to the CCSD(T)/CBS reference interaction energies $E_{R}$ (kcal/mol) and conservative 3tJ protocol data~\cite{Dubecky2013}, along with the corresponding differences $\Delta$ and statistics: mean error (ME, kcal/mol), mean unsigned error (MUE, kcal/mol) and relative unsigned error (RUE,\%), for each subset and the whole considered set (total), respectively.
}
\centering
\begin{tabular}{lc|cr|cr}
\hline
\hline
Complex          &$E_{R}$ &$E$/3tJ$^d$& $\Delta_{R}$ & $E$/2tJ$^c$ & $\Delta_{R}$\\
\hline
(teaching set)   &              &                  &        &                 &       \\
Ammonia dimer    &-3.15$^a$     &   \am            & -0.05  &  -3.21$\pm$0.09 & 0.06  \\
Water dimer      &-5.07$^a$     &   -5.26$\pm0.08$ &  0.19  &  -5.28$\pm$0.10 & 0.21  \\
Hydrogen fluoride dimer
                 &-4.62$^b$     &  -4.68$\pm0.10$  &  0.10  &  -4.77$\pm$0.12 & 0.19  \\
Methane dimer    &-0.53$^a$     &   -0.44$\pm0.05$ & -0.09  &  -0.60$\pm$0.07 & 0.07  \\
Ethene dimer     &-1.48$^a$     &   -1.47$\pm0.09$ & -0.01  &  -1.53$\pm$0.13 & 0.05  \\
Ethene/ethyne    &-1.50$^a$     &   -1.56$\pm0.08$ &  0.06  &  -1.54$\pm$0.14 & 0.04  \\\hline
                 &              &        ME:       &  0.033 &           ME:   & 0.104 \\
                 &              &        MUE:      &  0.083 &          MUE:   & 0.104 \\
                 &              &        RUE:      &  4.78  &          RUE:   & 4.94  \\
\hline
(test set)               &            &                &        &                  \\
Benzene/H$_2$            & -1.03$^c$  & -              &   -    & -1.08$\pm0.10$ & 0.04\\
Benzene/water            & -3.29$^a$  & -3.53$\pm0.13$ &  0.24  & -3.21$\pm0.15$ & -0.08\\
Benzene/methane          & -1.45$^a$  & -1.30$\pm0.13$ & -0.15  & -1.50$\pm0.15$ & 0.05\\
Benzene dimer T          & -2.71$^a$  & -2.88$\pm0.16$ &  0.17  & -2.53$\pm0.23$ & -0.18\\
Adenine-Thymine S        & -11.66$^a$ & -              &    -   &-11.00$\pm0.30$ & -0.66\\\hline
                         &            &                &        & ME:     &-0.167 \\
                         &            &                &        & MUE:    & 0.203 \\
                         &            &                &        & RUE:     & 4.44  \\                        
\hline\hline
\end{tabular}\\
\begin{tabular}{l}
$^a$ Takatani et al.~\cite{Takatani2010}, 
$^b$ CCSDT(Q), \v{R}ez\'a\v{c} et al.~\cite{Rezac2013},
$^c$ This work.
$^d$ Dubeck\'y et al.~\cite{Dubecky2013},
\end{tabular}
\label{tabres}
\end{table*}
\section{Benchmarks}
The production FN-DMC results obtained by the scheme involving simplified Jastrow term (2tJ) are reported and compared to the CCSD(T)/CBS and more traditional  protocol (3tJ) data in~Tab.~\ref{tabres}. 
The mean error (ME) and the mean unsigned error (MUE) of the 2tJ scheme, with respect to the CCSD(T)/CBS reference data, are both found to be about \mbox{0.1~kcal/mol} respectively and the mean relative unsigned error (RUE) reaches 4.9\%. The maximum interaction energy deviations with respect to the reference reach \mbox{$~\sim0.2$~kcal/mol}, observed in hydrogen-bonded water and hydrogen-fluoride dimers.
For comparison, the MUE of the FN-DMC results obtained with the 3tJ scheme amounts to 0.08 kcal/mol (RUE: 4.8~\%) and the maximum deviation of
\mbox{$~\approx$ 0.2~kcal/mol} is obtained only in the case of water dimer. This may be further improved by the distinct Jastrow factor (cf.~Tab.~\ref{tabrescmp})~\cite{Dubecky2013} if required, as already mentioned, nevertheless here we are interested in a computationally less expensive computations - 2tJ scheme. The remaining 2tJ values differ by no more than \mbox{0.1~kcal/mol} from the CCSD(T)/CBS reference. 

The favourable performance of the 2tJ scheme is further demonstrated when it comes to larger complexes, where MUE amounts only to \mbox{0.2}~kcal/mol \mbox{(RUE: 4.4\%)}.
Overall, the data in ~Tab.~\ref{tabres} clearly show that the faster 2tJ scheme is able to attain benchmark results, close to the subchemical accuracy in most systems.
These results also presumably indicate, that efficient FN-error cancellation~\cite{Dubecky2013} takes place in the considered set of complexes. On the other hand, of the 2tJ would be used in larger complexes, a possible error-accumulation is possible in principle, nevertheless the concept of chemical accuracy in these cases looses its importance since the absolute value of the interaction energy grows and what should be taken as as measure of reliability of any theoretical approach is the relative energy deviation. The relative error of our approach does not noticeably increase with the size of the system which gives us confidence for use in large complexes. An exhaustive assessment of a general applicability and limits of the presented methods is a next goal of our work.

We note in passing that in the case of the benzene/H$_2$ complex, we report our own CCSD(T)/CBS reference value of \mbox{-1.033~kcal/mol}~(Tab.~\ref{tabres}), obtained by a standard CBS extrapolation/correction technique (cf. the Methods) which improves upon the existing value of \mbox{-1.037~kcal/mol}~\cite{Rubes2009}. The obtained QMC value (-1.08$\pm$0.1~kcal/mol) agrees (within the error bar) with the reference, as well as with the reported value of 0.96$\pm$0.08~kcal/mol~\cite{Beaudet2008} obtained at a slightly different geometry. 

{Finally, we have extensively tested the new 2tJ protocol on a set of 24 noncovalent complexes (A24)~\cite{Rezac2013}. The results are summarized in the Tab.~\ref{tabresA24}.}
\begin{table*}[t!]
\scriptsize
\caption{The FN-DMC interaction energies $E$ (kcal/mol) of the A24 set, obtained by the 2tJ protocol, compared to the CCSD(T)/CBS reference interaction energies $E_{R}$ (kcal/mol) along with the corresponding differences $\Delta$, illustrations~\cite{PyMOL} and statistics: mean error (ME, kcal/mol), mean unsigned error (MUE, kcal/mol) and relative unsigned error (RUE,\%).}
\newcolumntype{C}{>{\centering\arraybackslash} m{1.5cm} }
\centering
\begin{tabular}{cCm{1.6cm}ccr||cCm{1.6cm}ccr}
\hline
\hline
Label&Complex & & $E_R$ & $E$/2tJ & $\Delta$ & Label&Complex & & $E_R$ & $E$/2tJ & $\Delta$  \\
\hline
&&&&&&&\\
1 & Water ammonia  & \includegraphics[width=50pt]{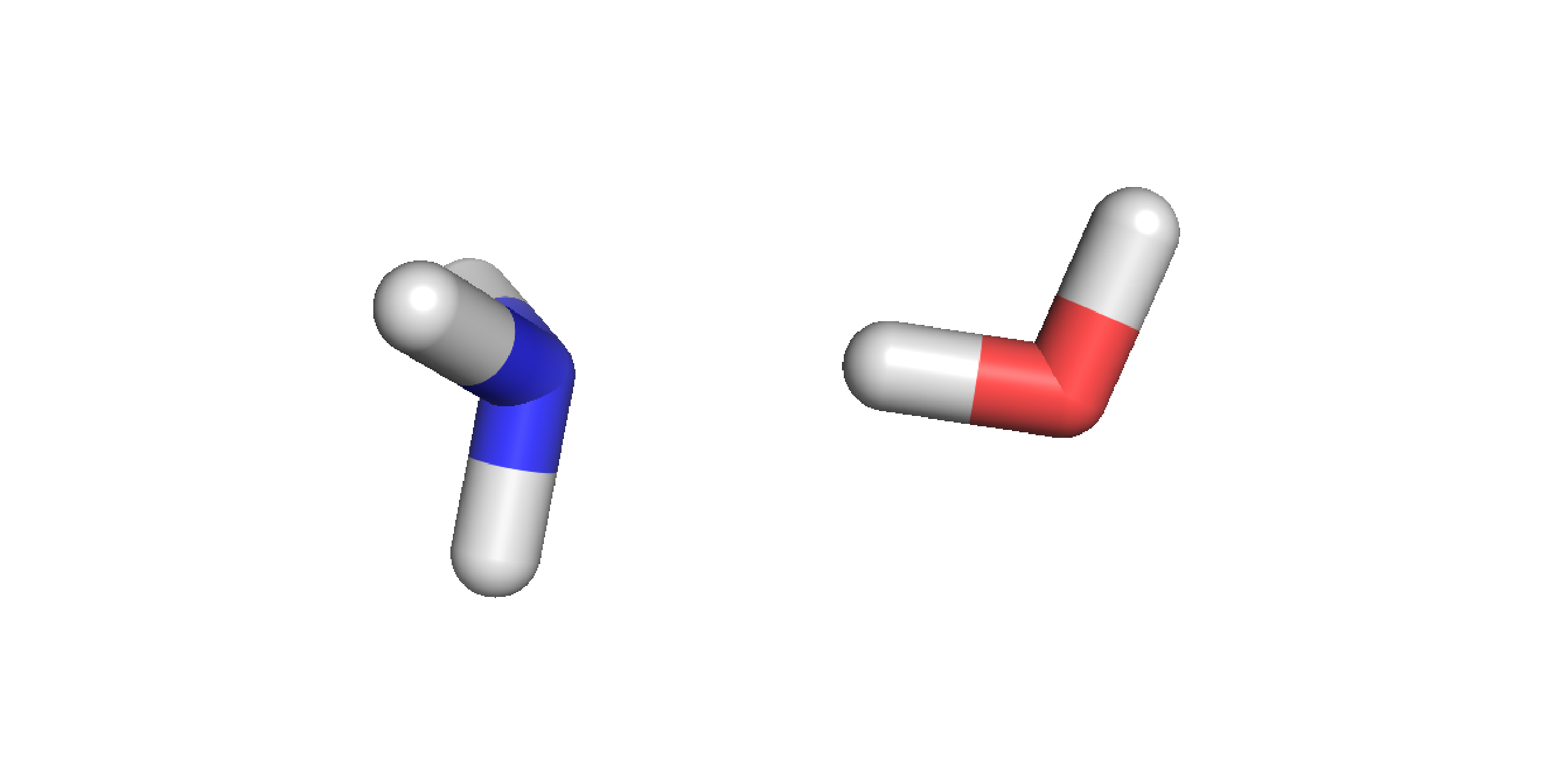}   &-6.493 & -6.71$\pm$0.07 & -0.22 & 13 & Ethene ammonia & \includegraphics[width=50pt]{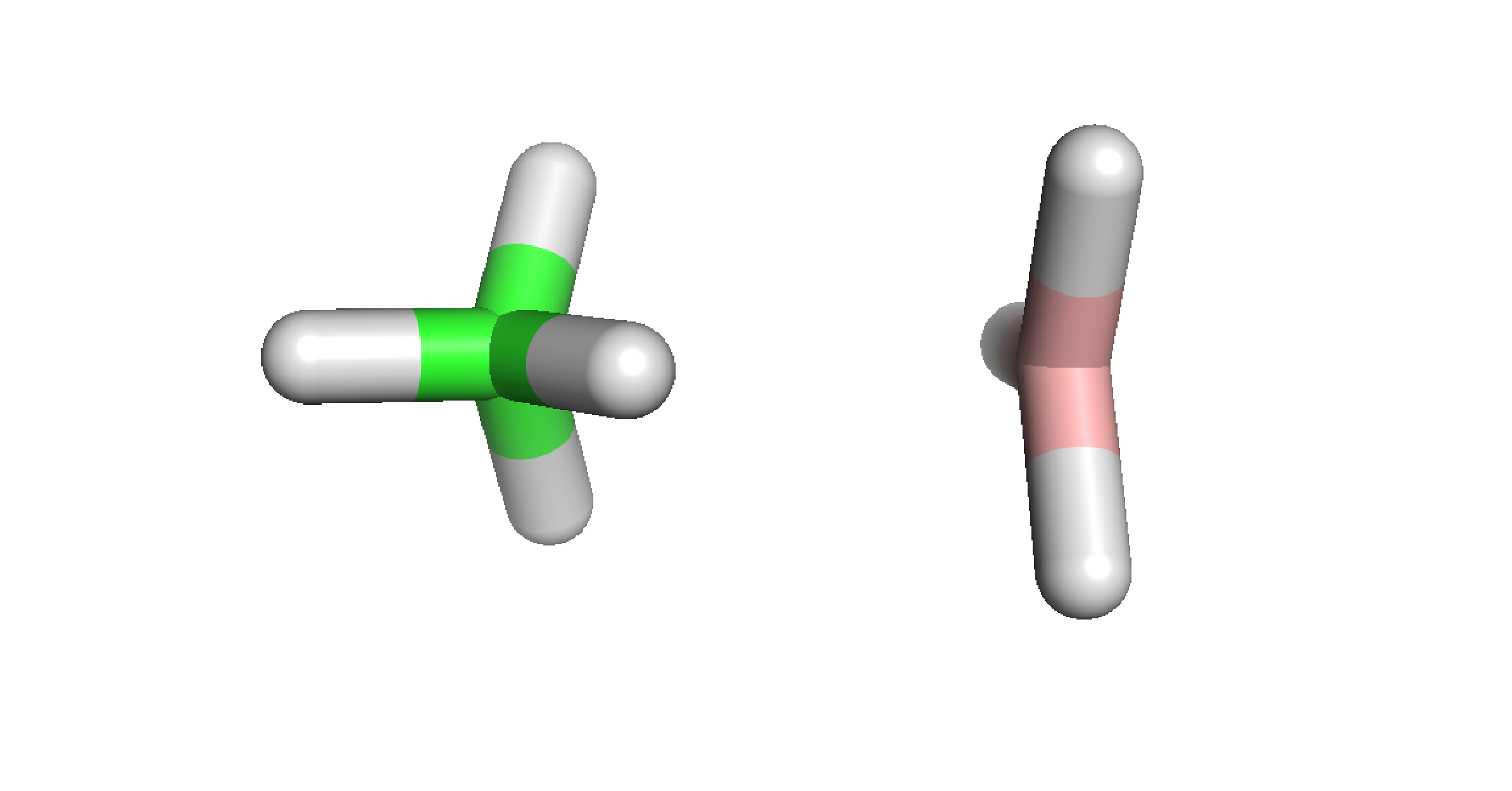}   & -1.374& -1.45$\pm$0.07 & -0.07 \\
2 & Water dimer    & \includegraphics[width=50pt]{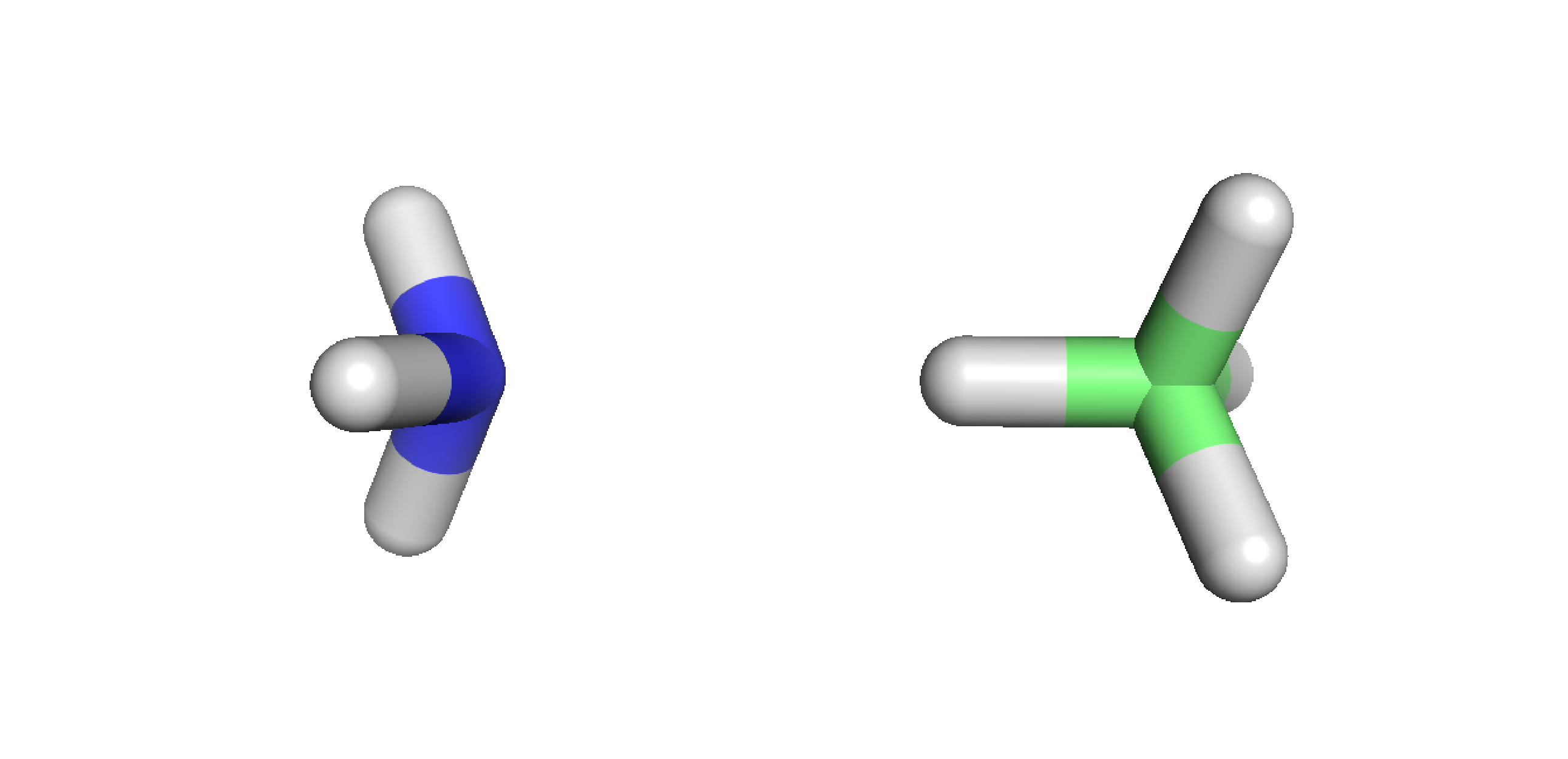}   &-5.006 & -5.30$\pm$0.05 & -0.29 & 14 & Ethene dimer   & \includegraphics[width=50pt]{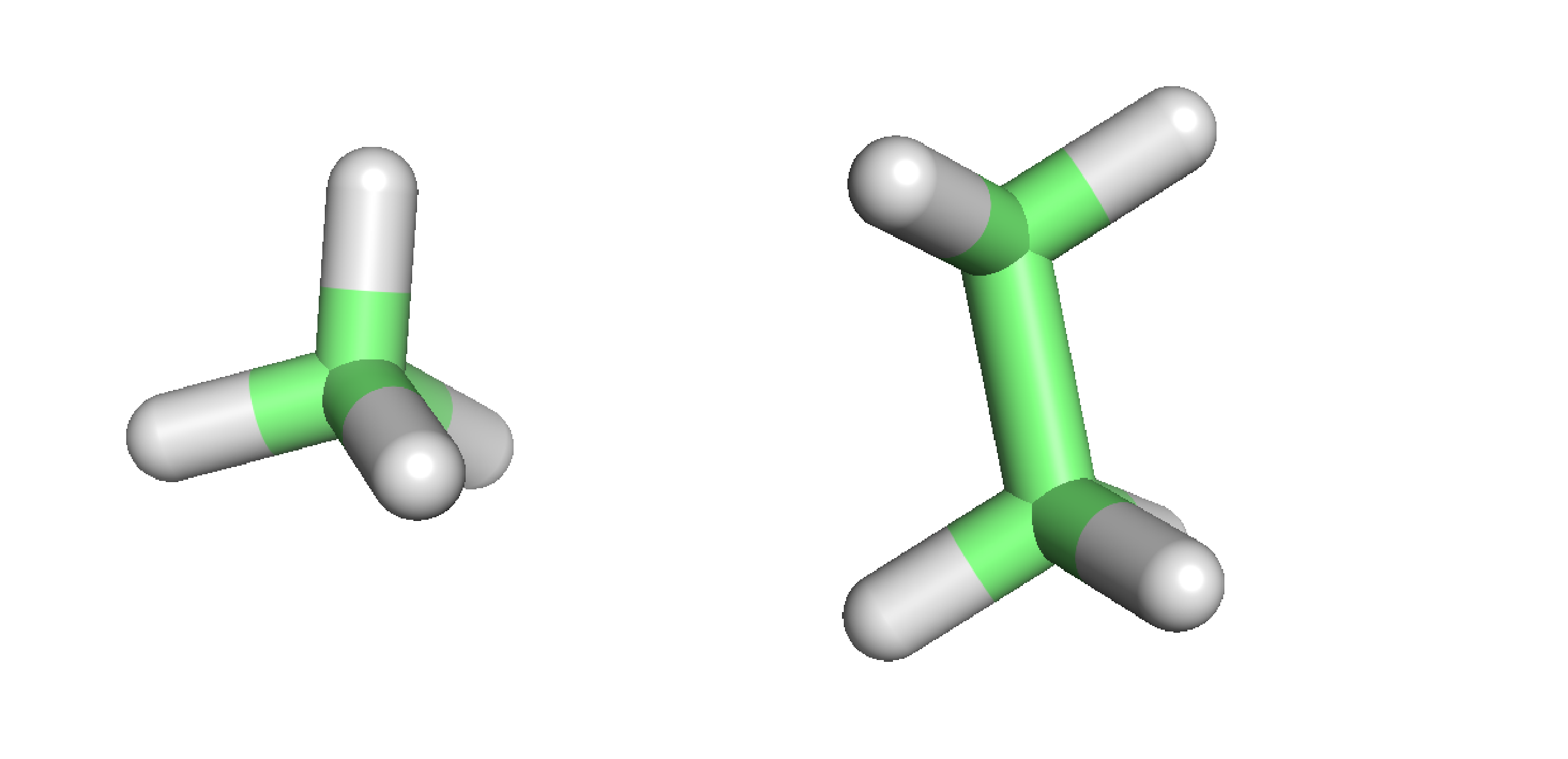}   & -1.09 & -1.08$\pm$0.07 & 0.01 \\
3 & HCN dimer      & \includegraphics[width=50pt]{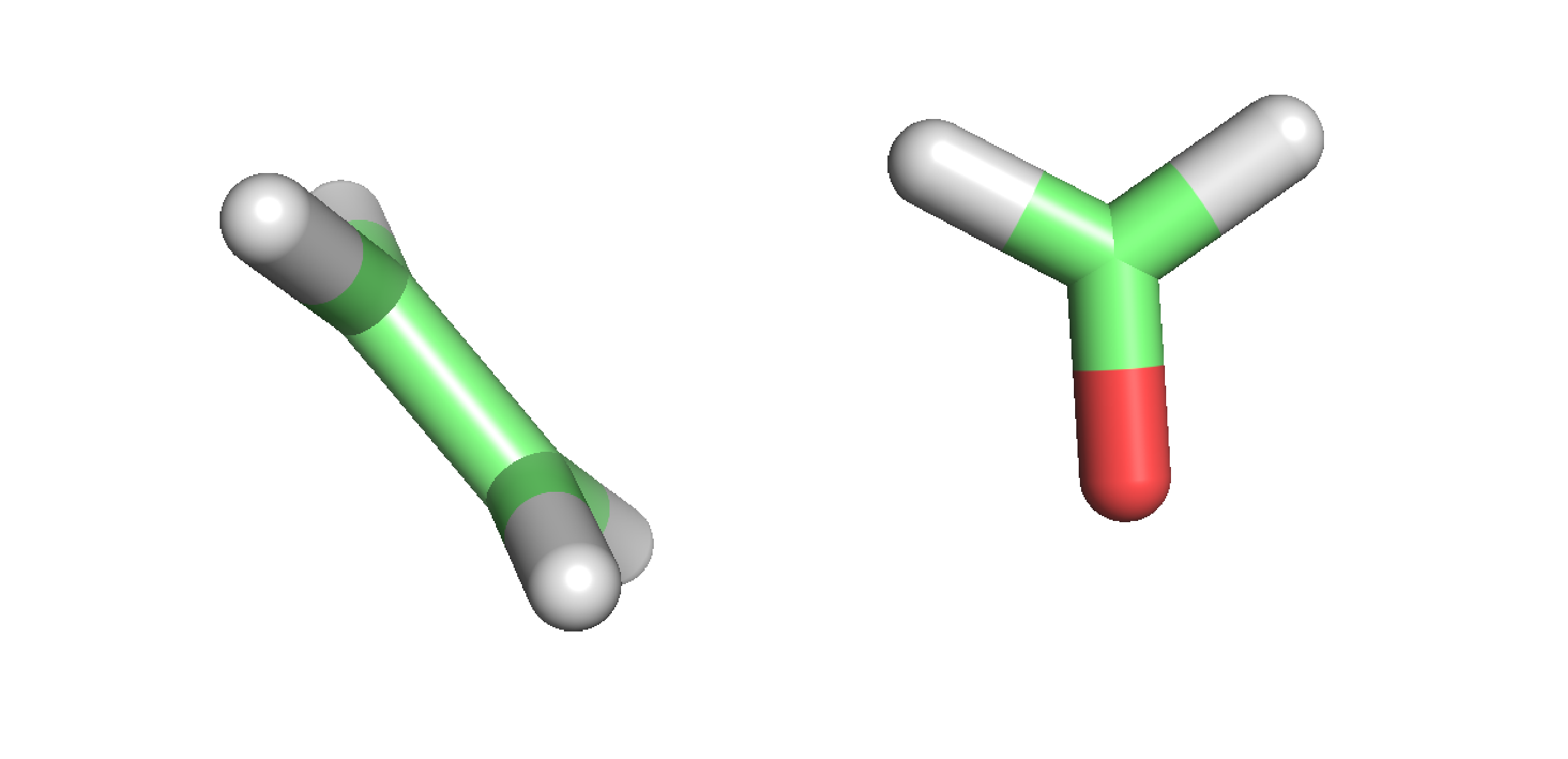}   &-4.745 & -5.09$\pm$0.08 & -0.35 & 15 & Methane ethene & \includegraphics[width=50pt]{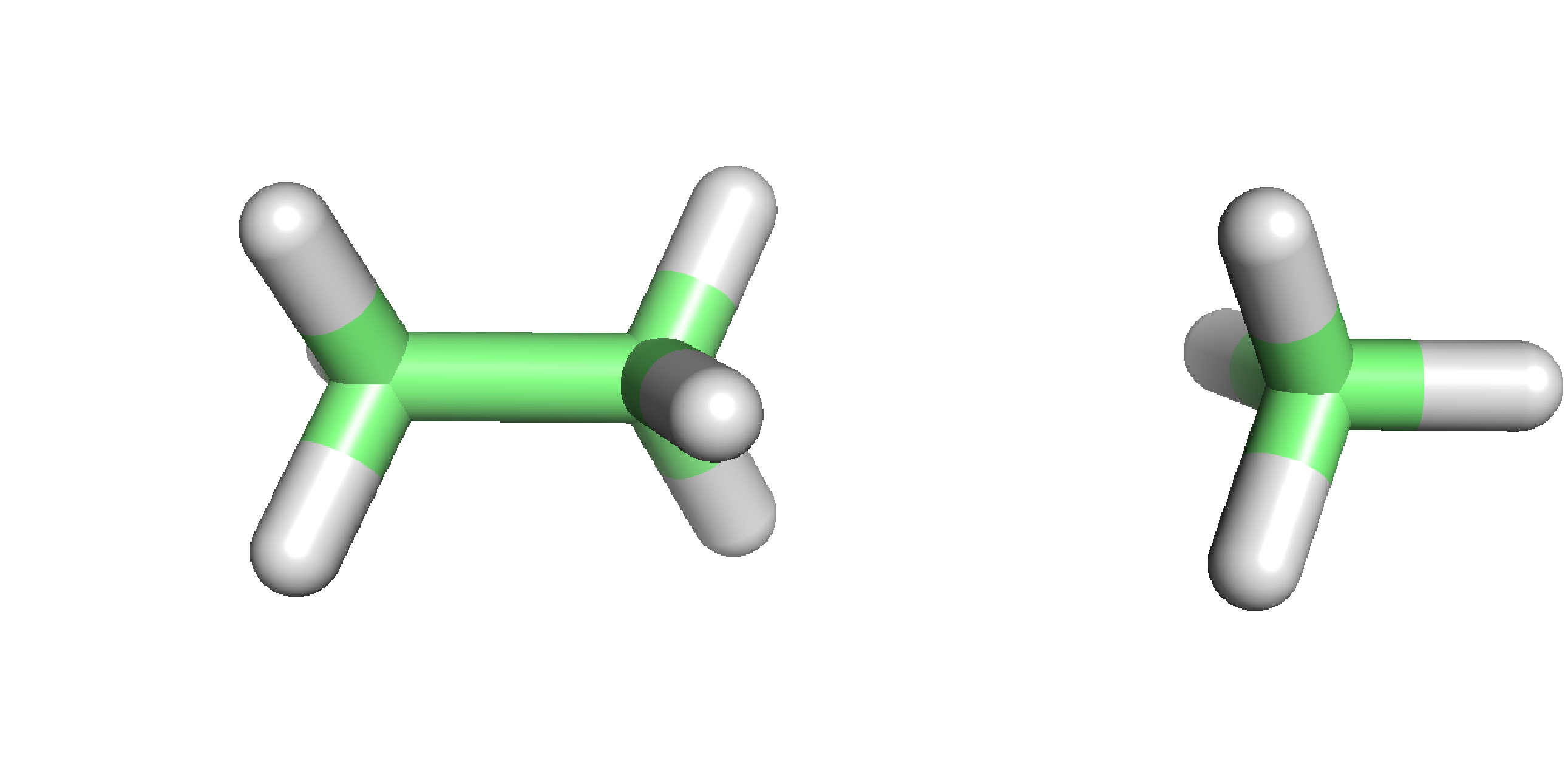}   & -0.502& -0.54$\pm$0.06 & -0.04 \\
4 & HF dimer       & \includegraphics[width=50pt]{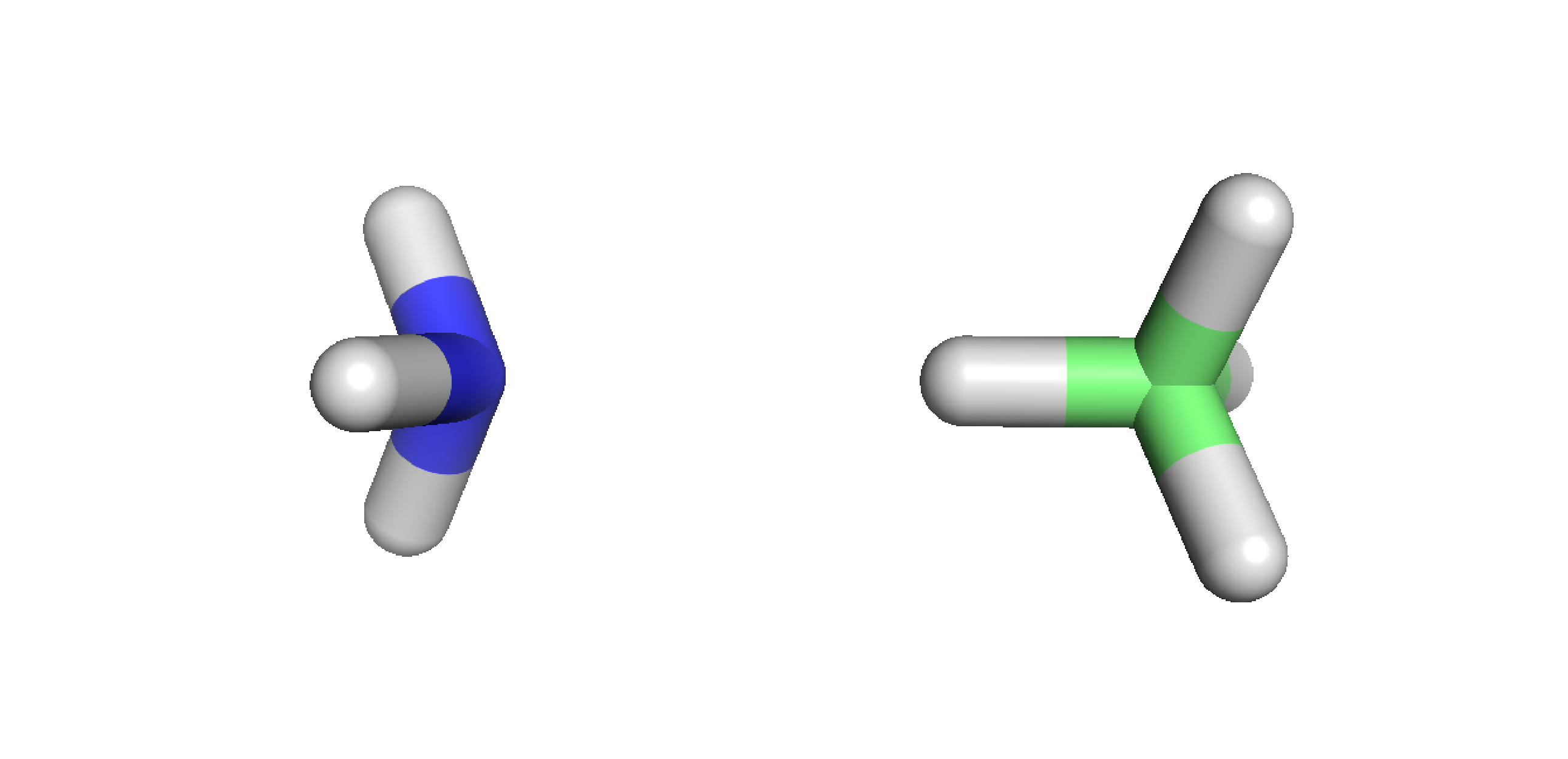}   &-4.581 & -4.88$\pm$0.05 & -0.30 & 16 & Borane methane & \includegraphics[width=50pt]{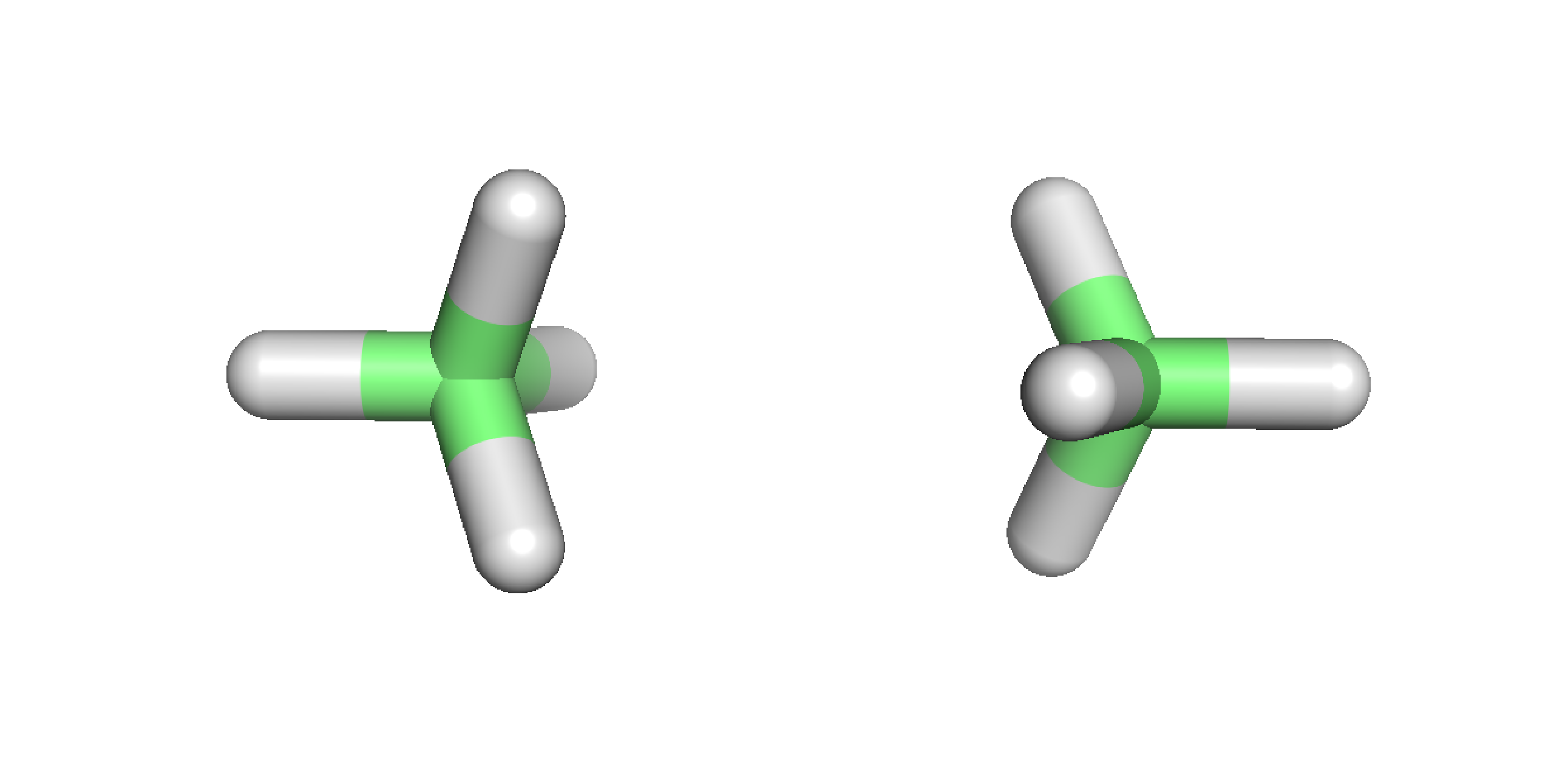}   & -1.485& -1.48$\pm$0.04 & 0.01 \\
5 & Ammonia dimer  & \includegraphics[width=50pt]{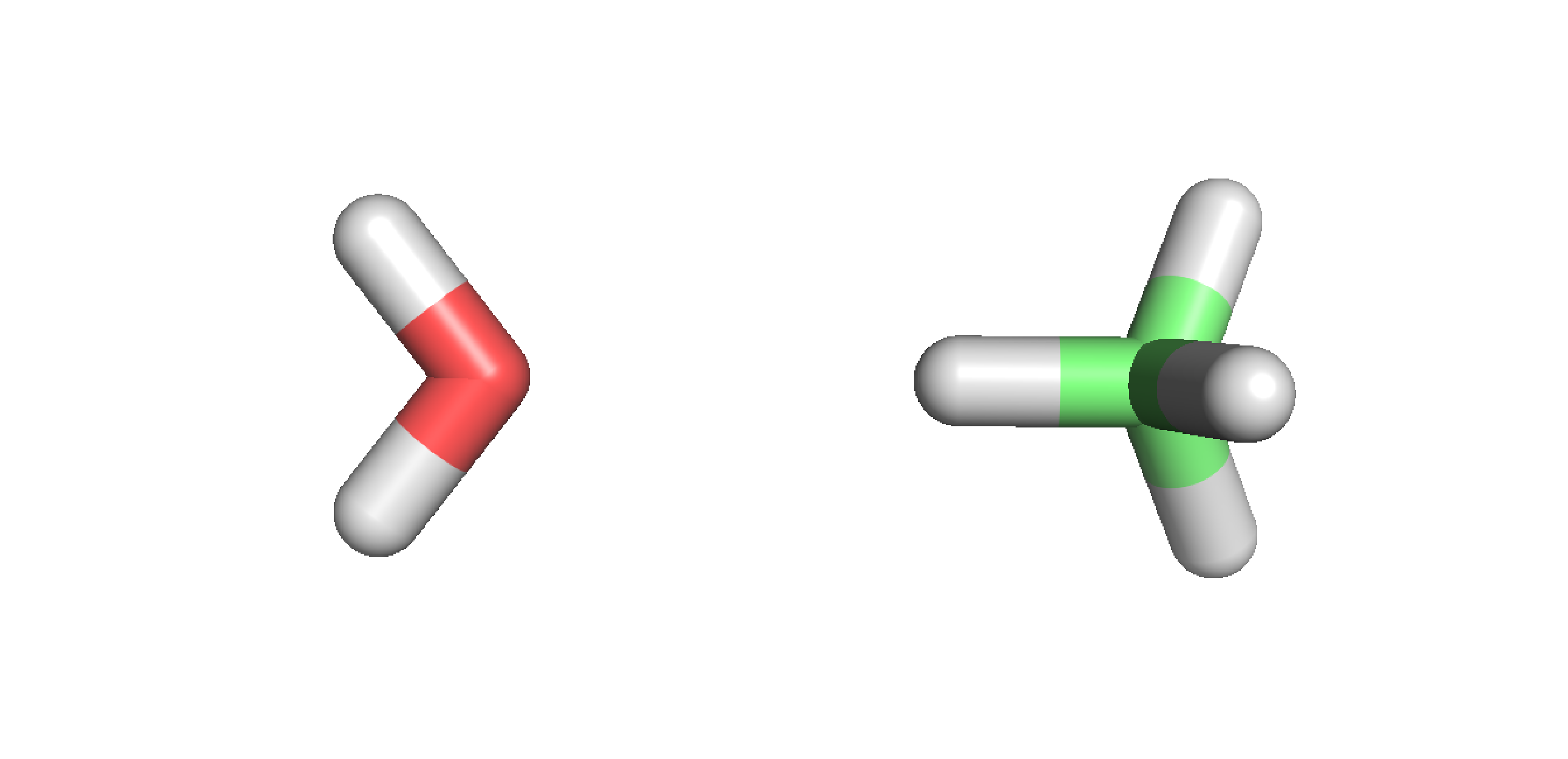}   &-3.137 & -3.30$\pm$0.04 & -0.17 & 17 & Methane ethane & \includegraphics[width=50pt]{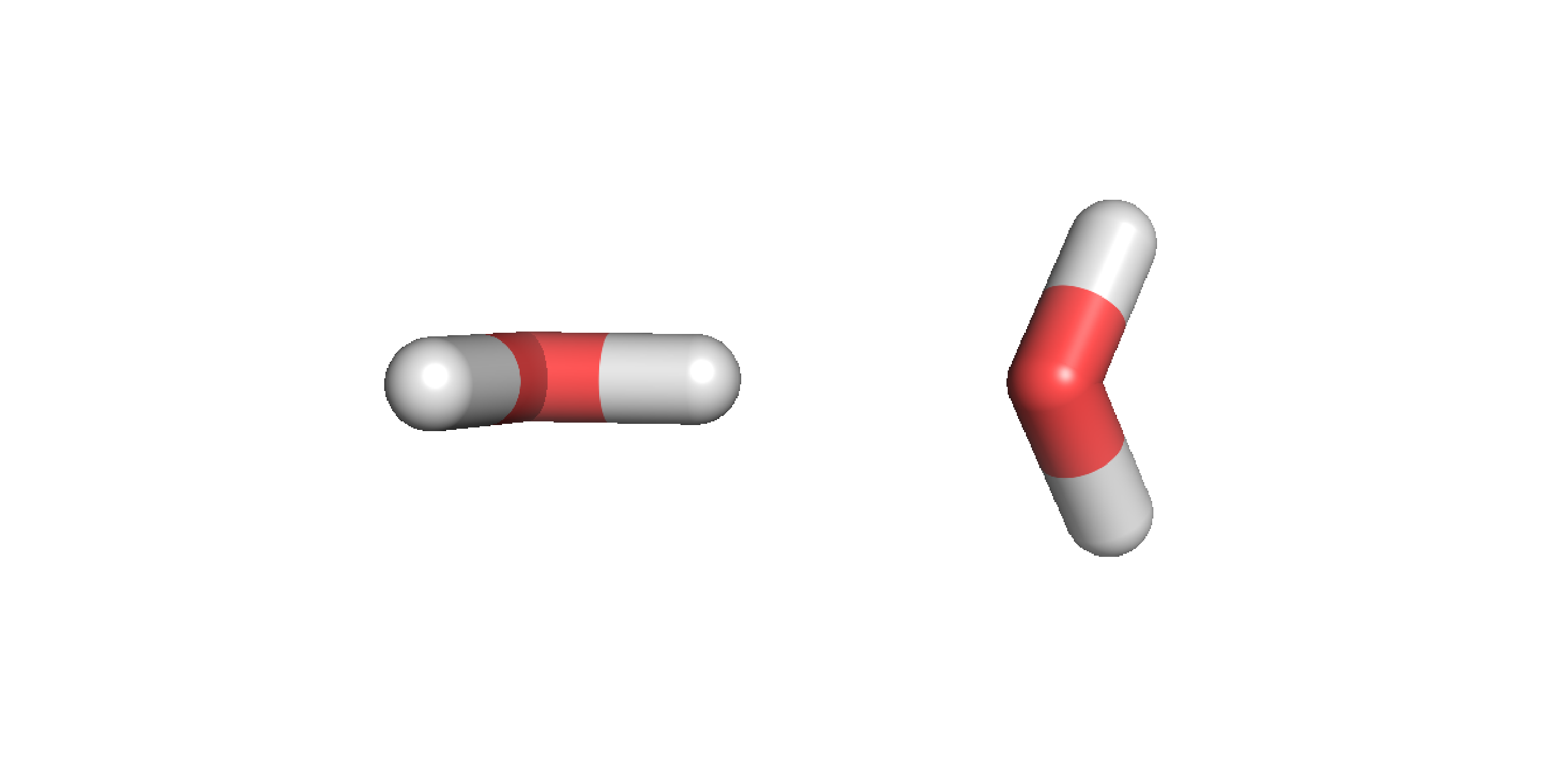}   & -0.827& -0.85$\pm$0.06 & -0.03 \\
6 & Methane HF     & \includegraphics[width=50pt]{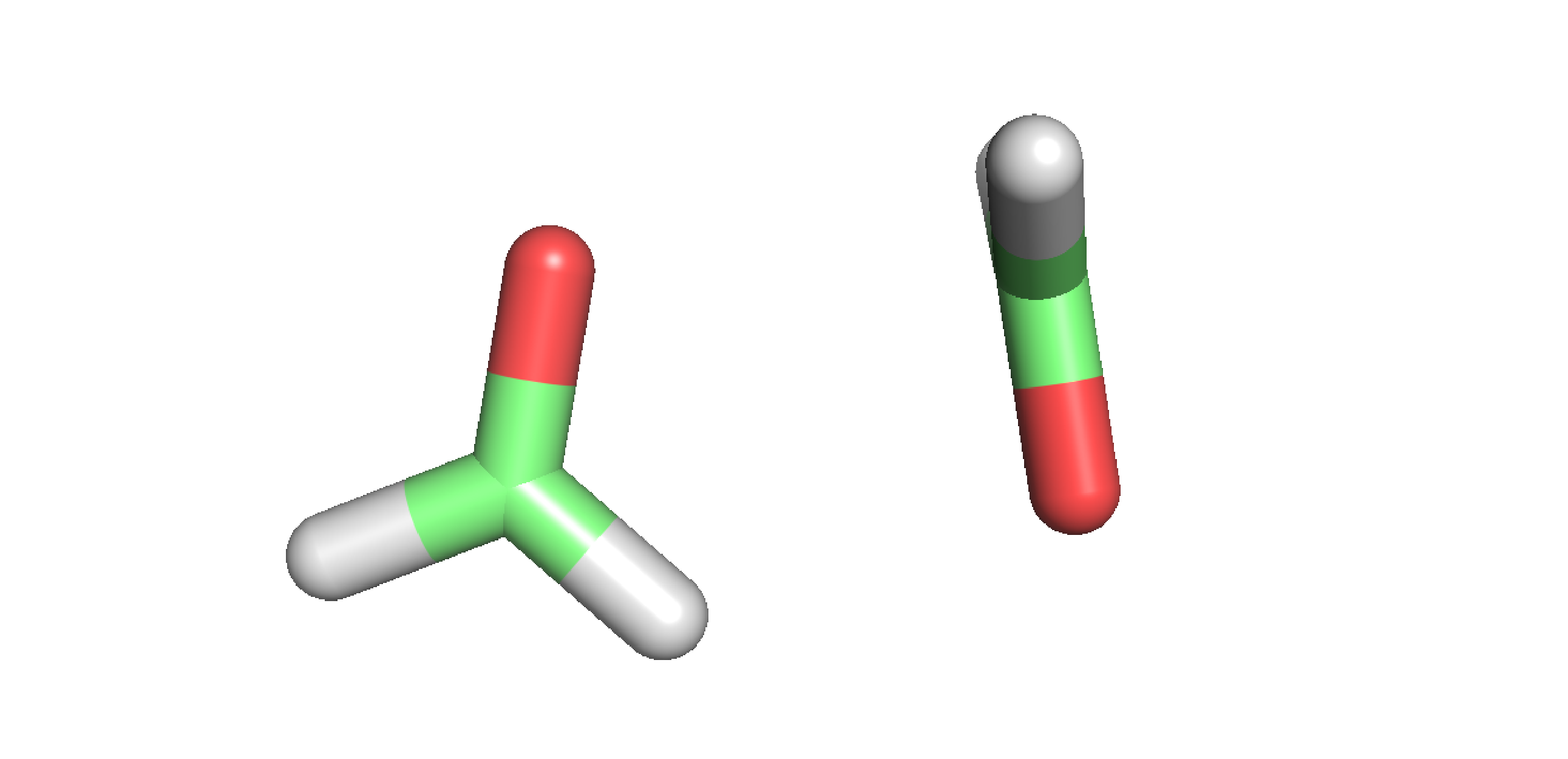}   &-1.654 & -1.29$\pm$0.07 & 0.37 & 18 & Methane ethane & \includegraphics[width=50pt]{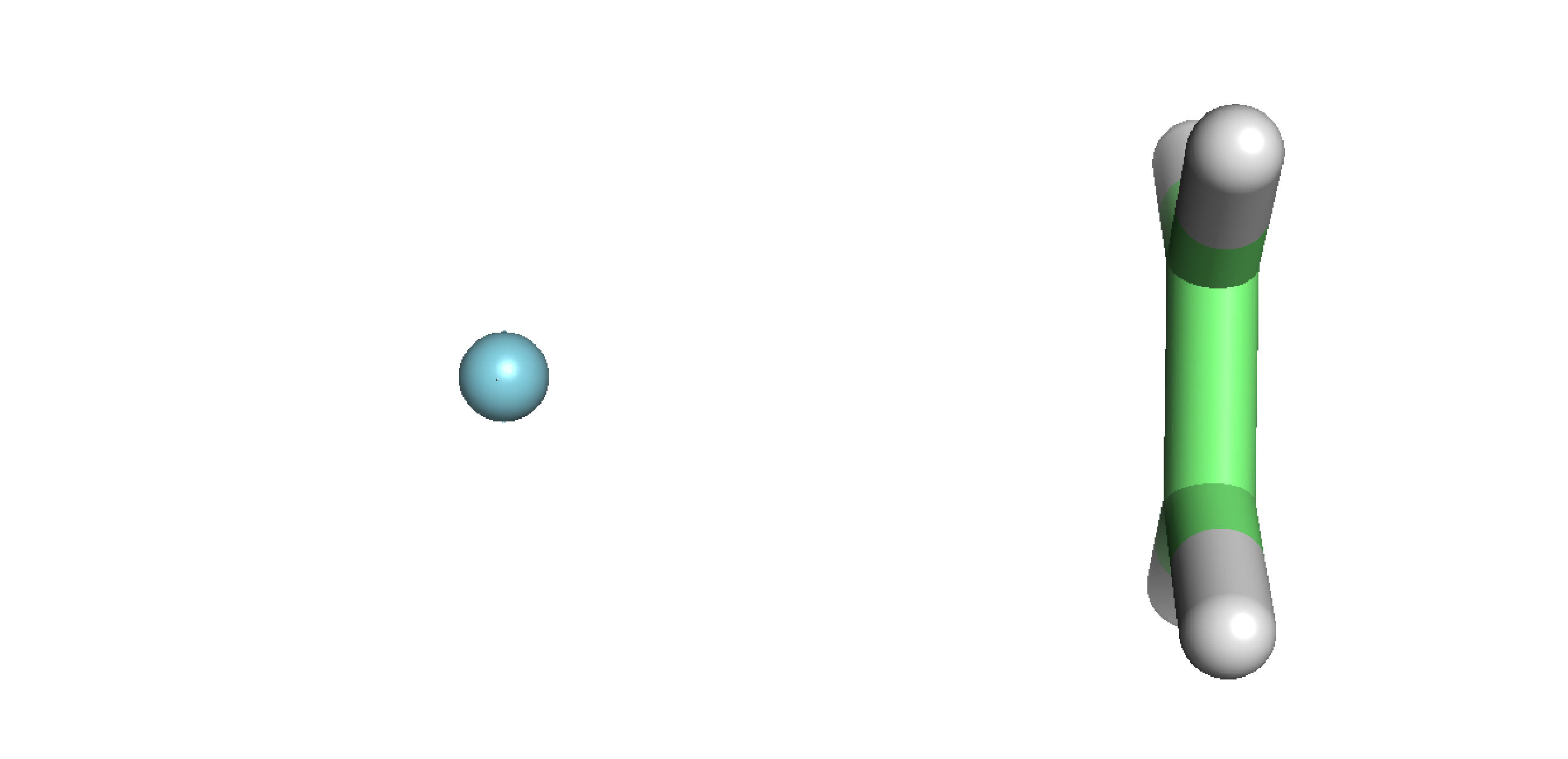}   & -0.607& -0.68$\pm$0.06  & -0.07 \\
7 & Ammonia methane& \includegraphics[width=50pt]{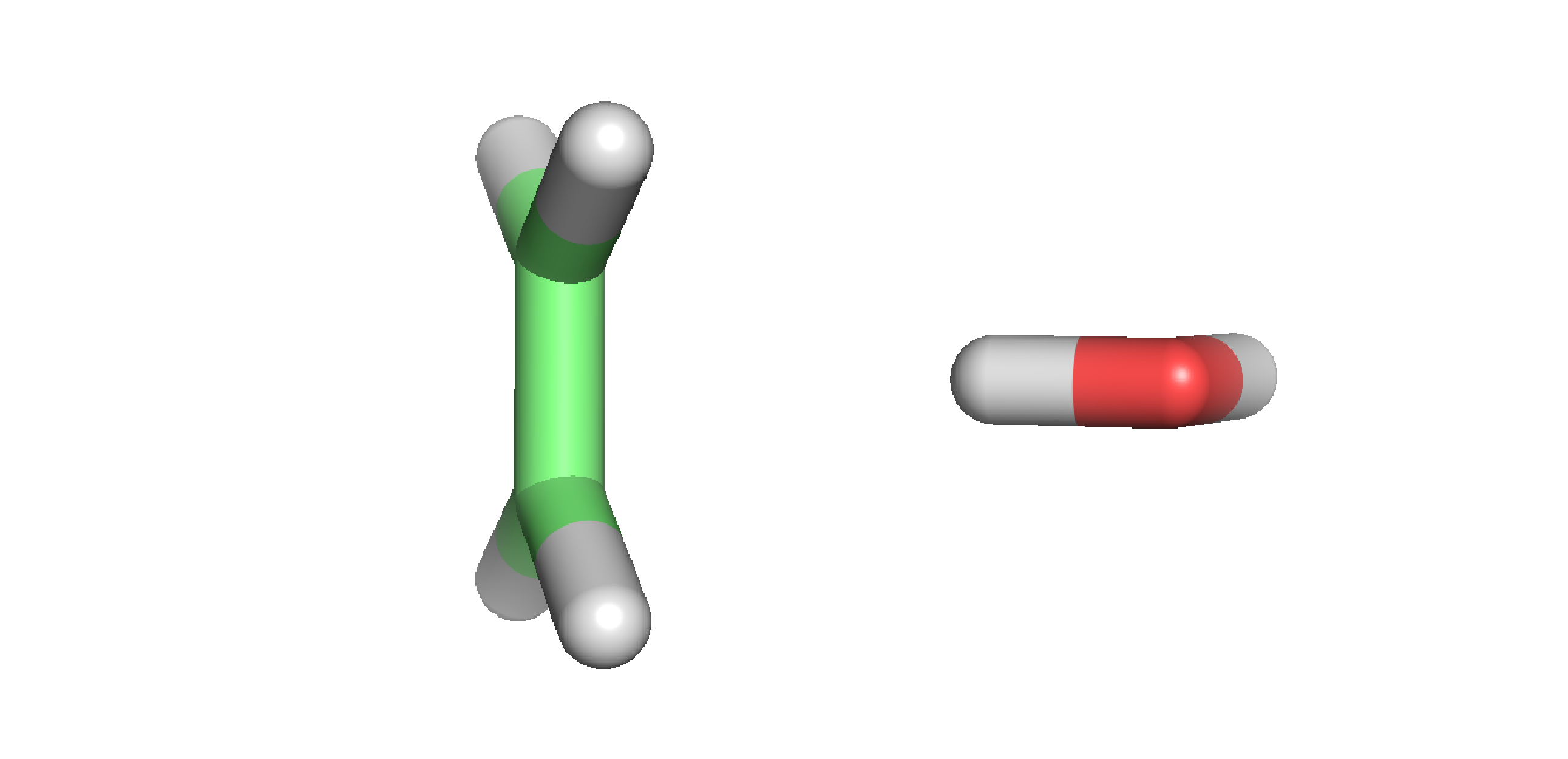}   &-0.765 & -0.83$\pm$0.06 & -0.06 & 19 & Methane dimer  & \includegraphics[width=50pt]{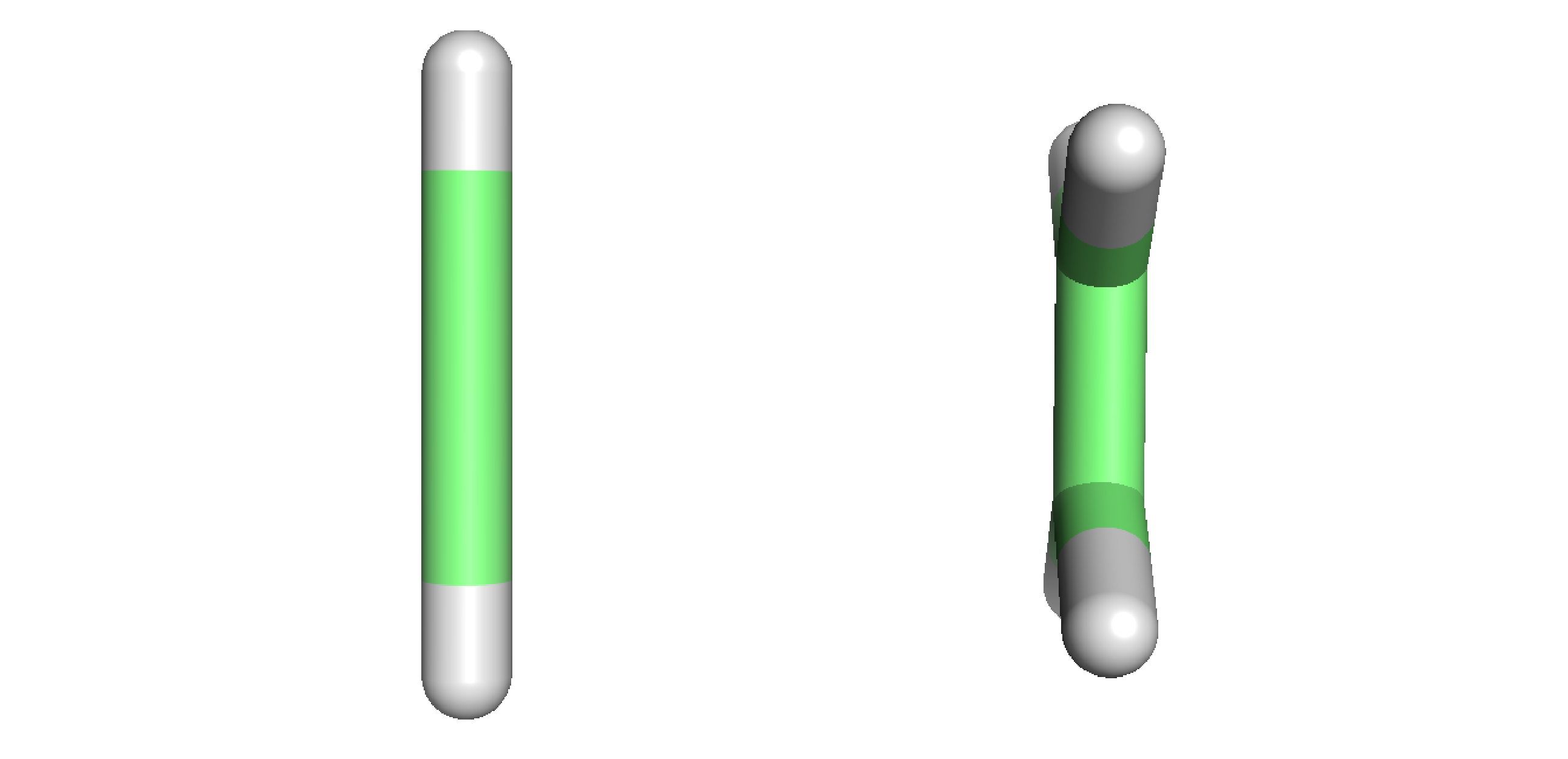}   & -0.533& -0.63$\pm$0.03 & -0.10 \\
8 & Methane water  & \includegraphics[width=50pt]{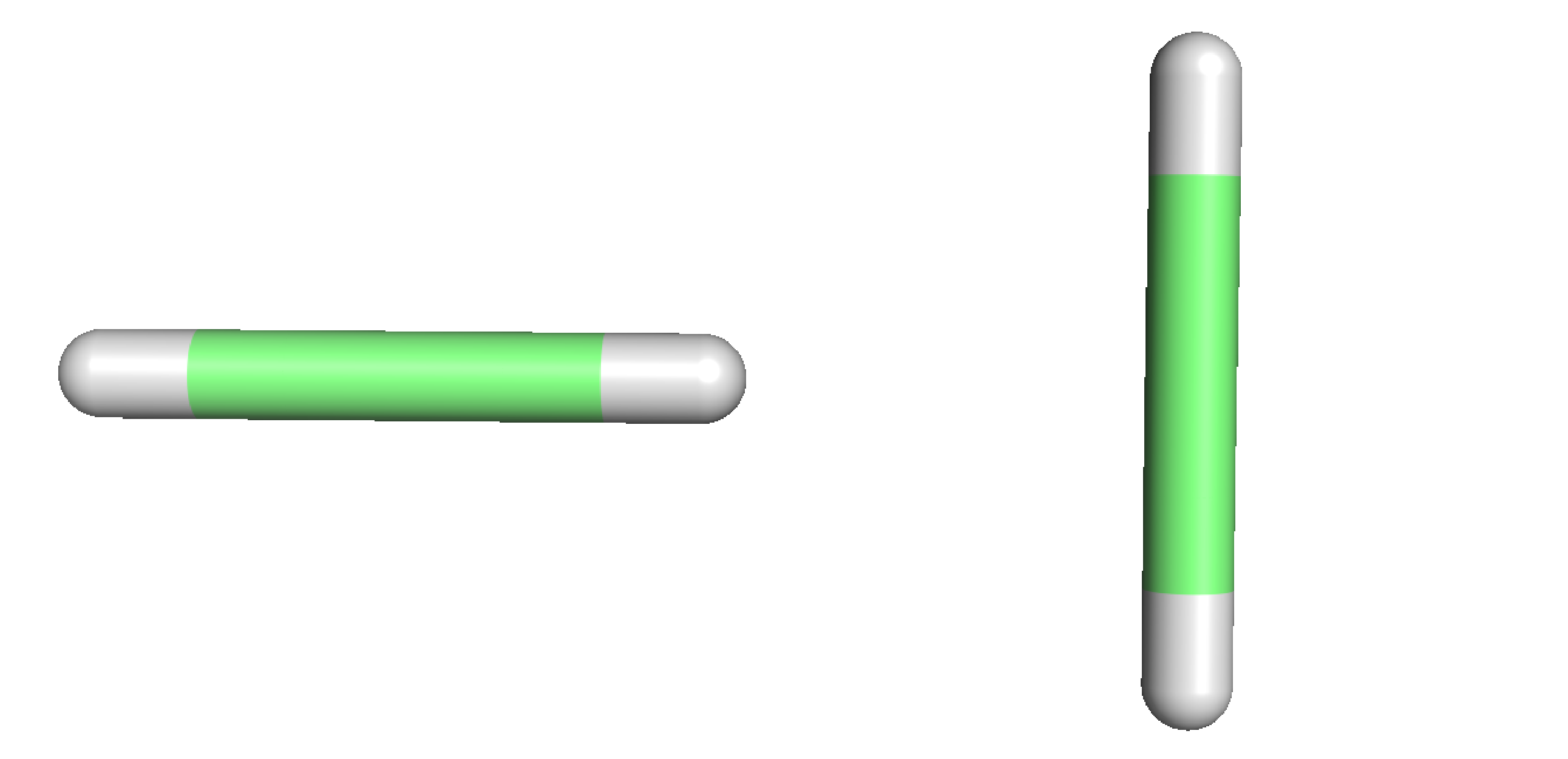}   &-0.663 & -0.57$\pm$0.06 & 0.09 & 20 & Ar methane     & \includegraphics[width=50pt]{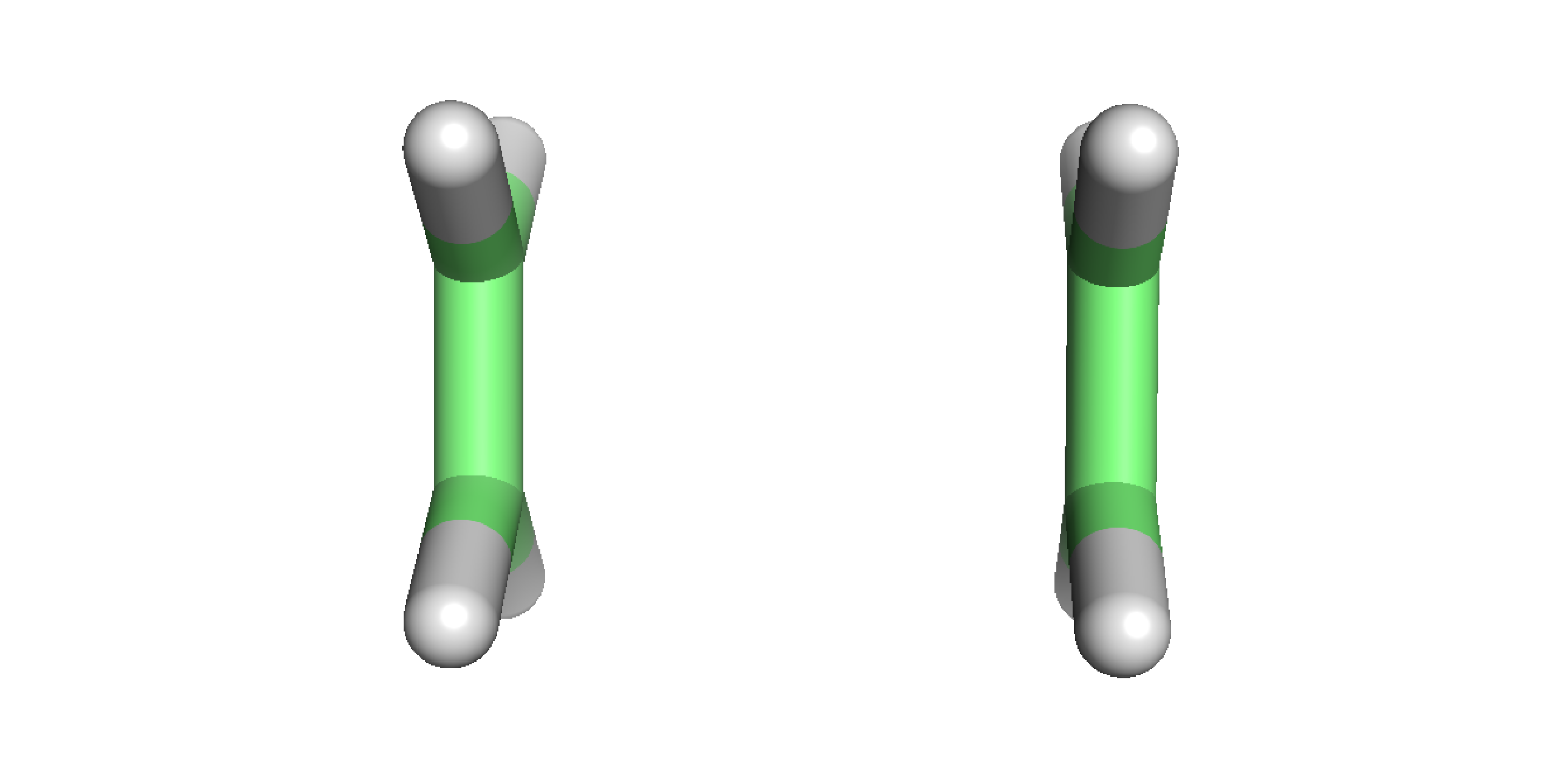}   & -0.405& -0.18$\pm$0.07  & 0.23 \\
9 & Formaldehyde dimer 
                   & \includegraphics[width=50pt]{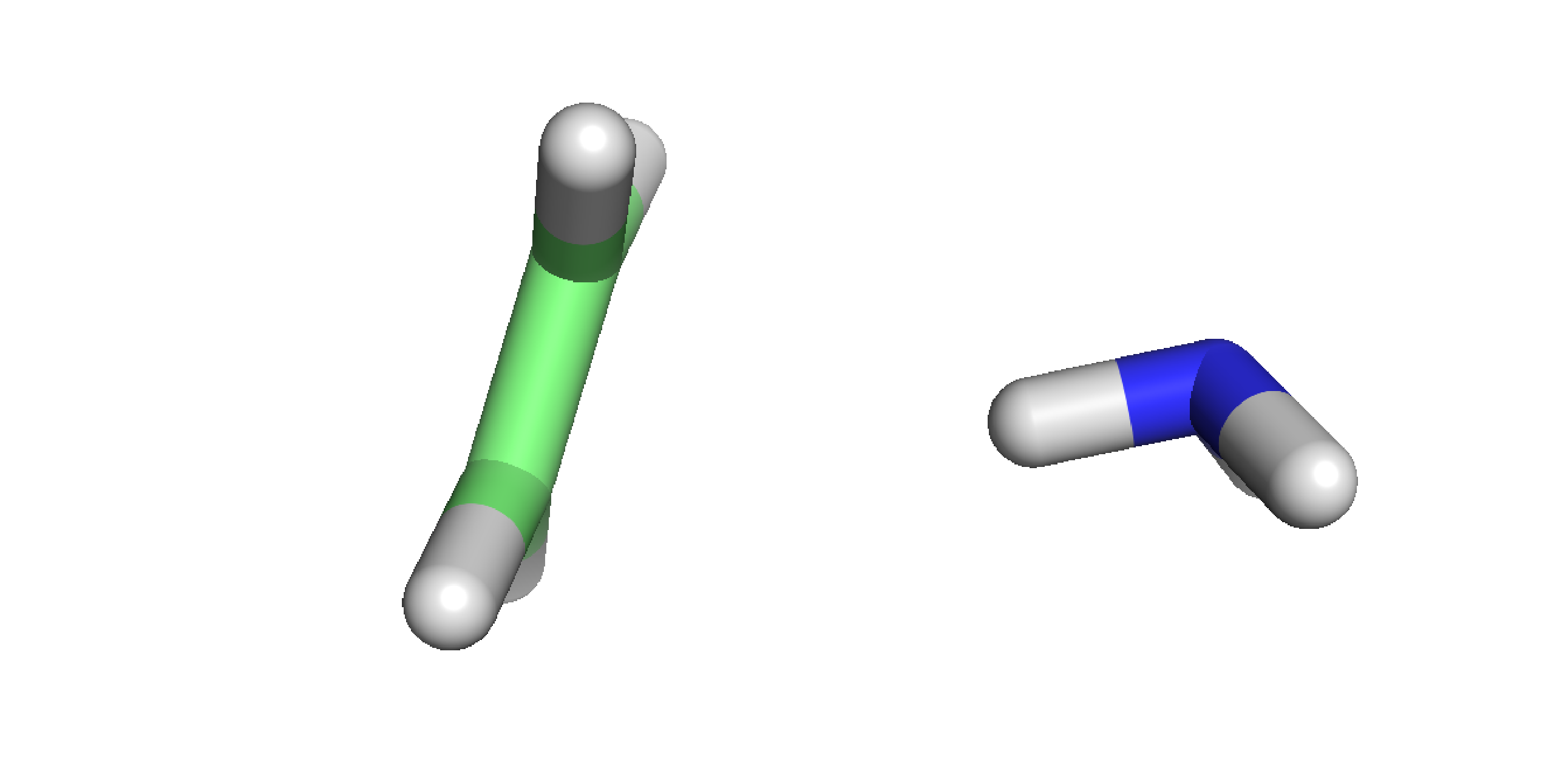}   &-4.554 & {-4.90}$\pm$0.10 & {-0.35}  & 21 & Ar ethene      & \includegraphics[width=50pt]{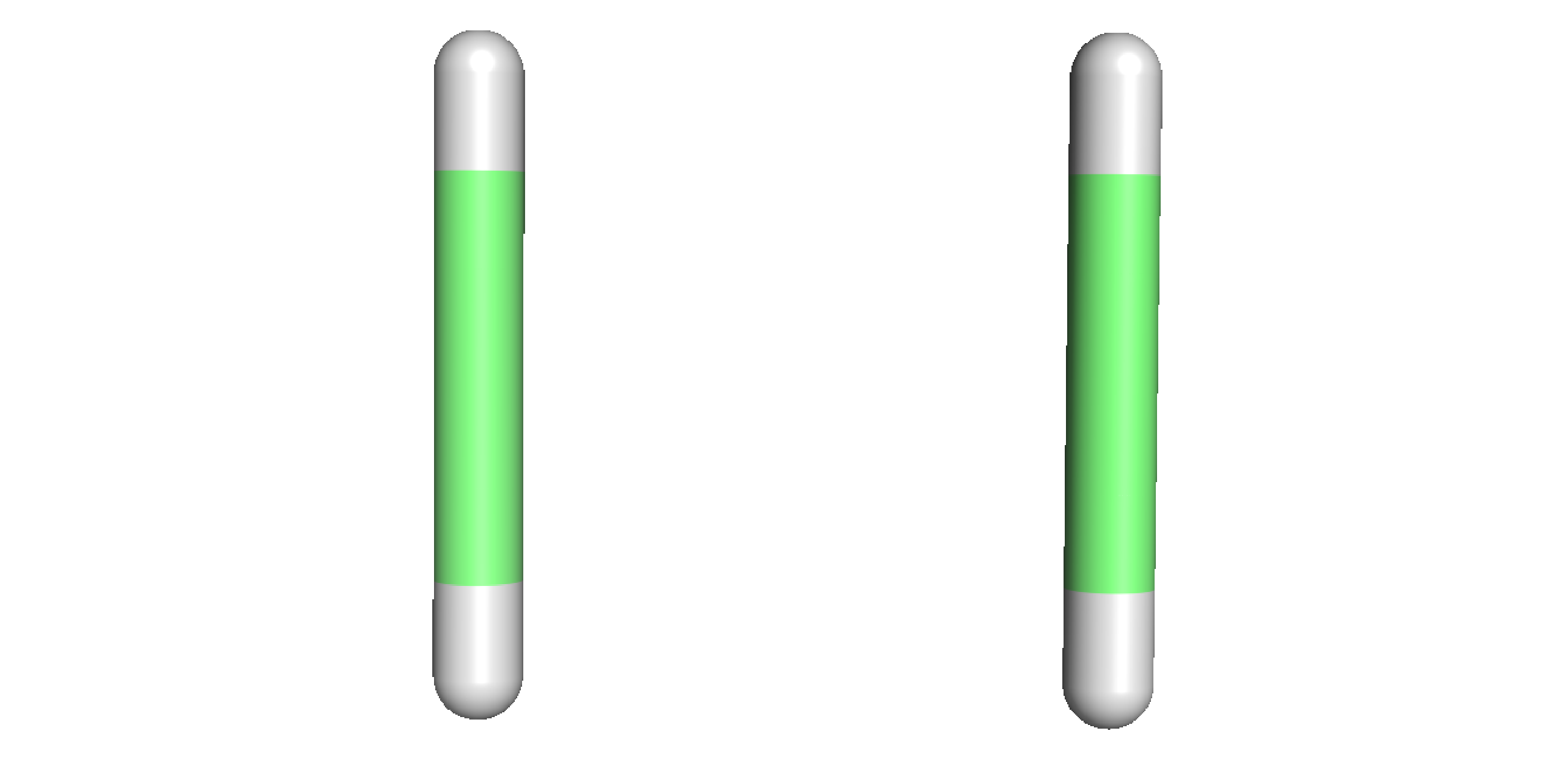}   & -0.364& -0.19$\pm$0.07 & 0.18 \\
10 & Water ethene  & \includegraphics[width=50pt]{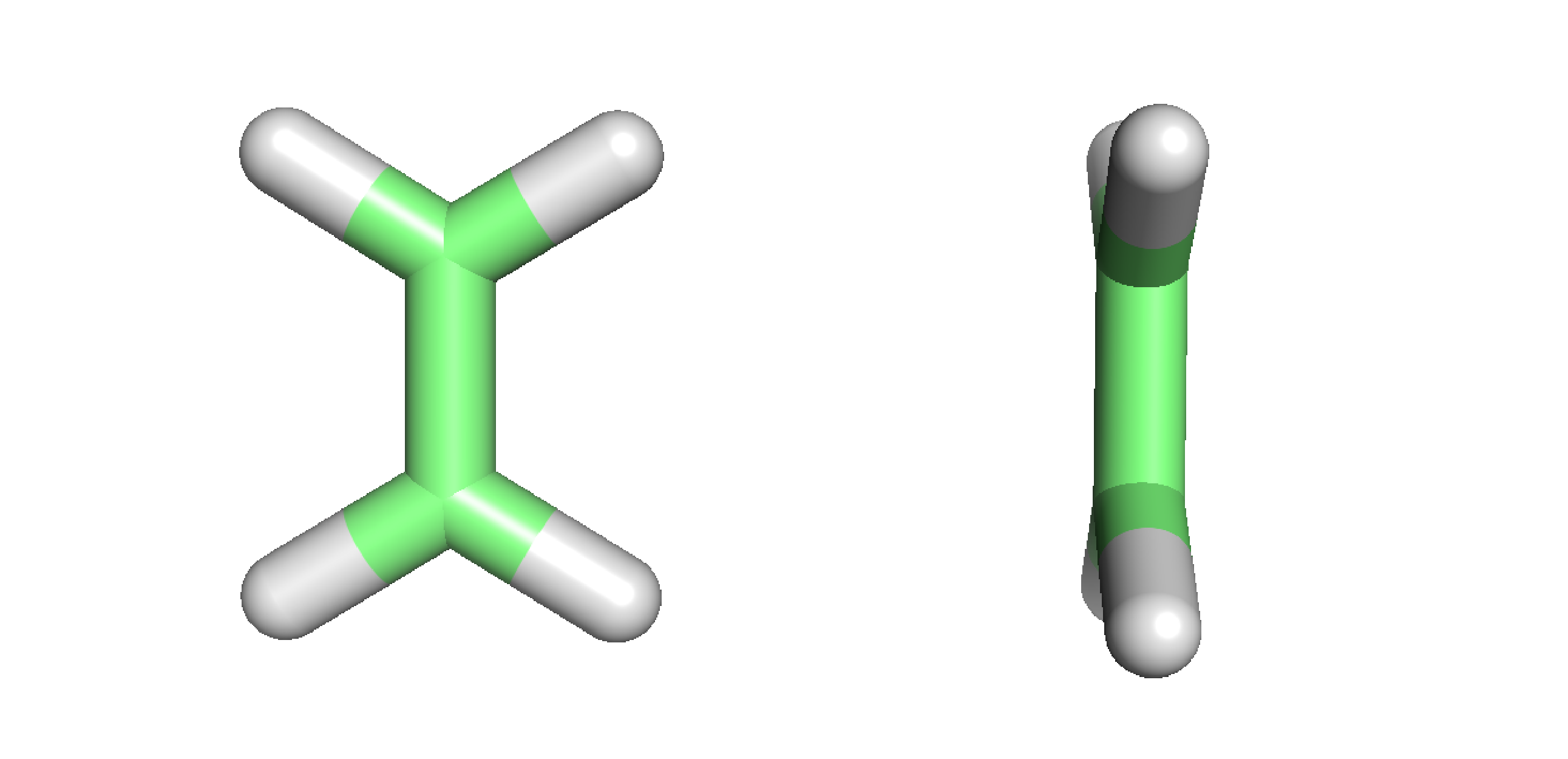}   &-2.557 & -2.55$\pm$0.07 & 0.01 & 22 & Ethene ethyne  & \includegraphics[width=50pt]{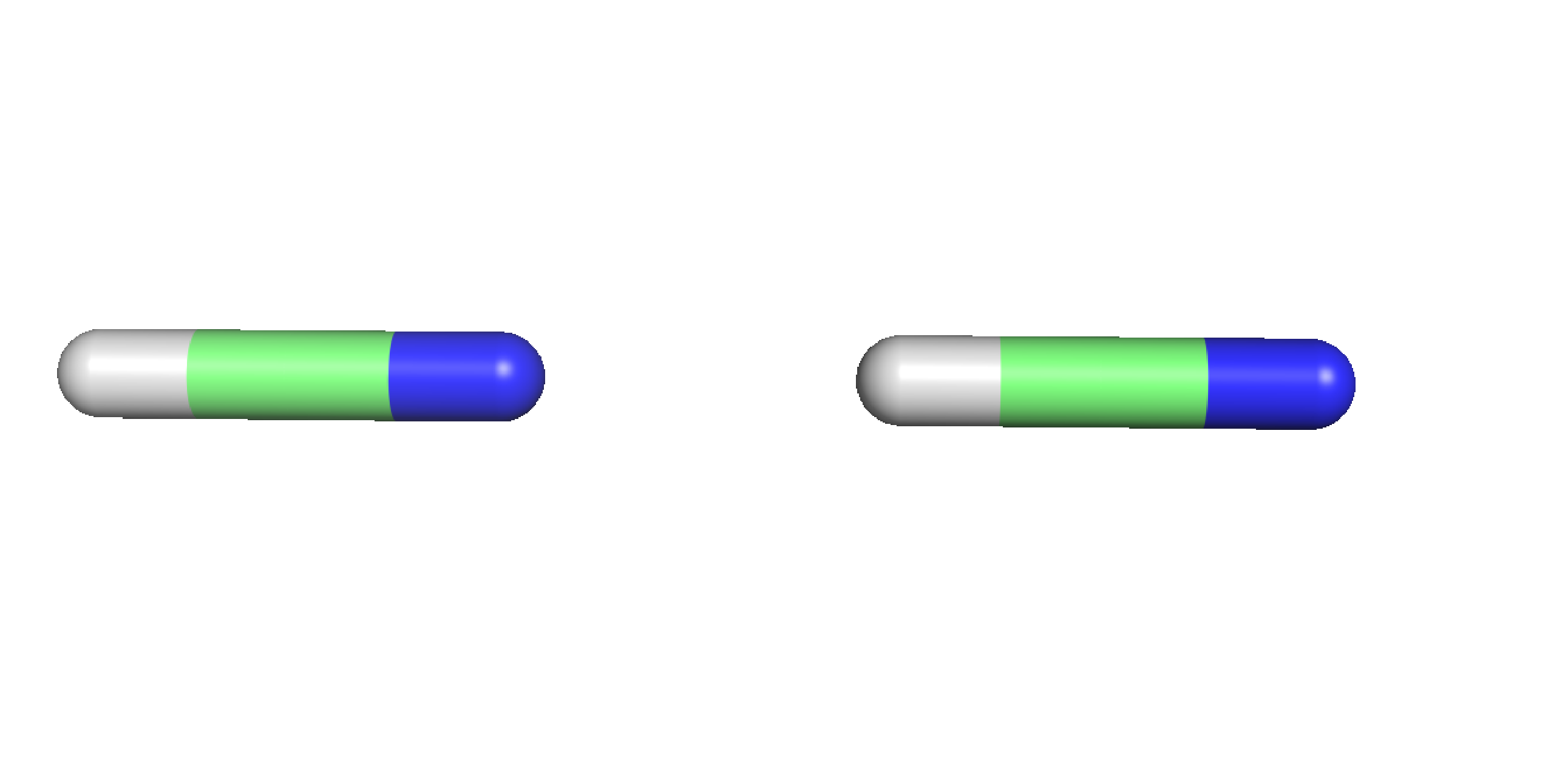}   & 0.821 &  0.94$\pm$0.07   & 0.12 \\
11 & Formaldehyde ethene
                   & \includegraphics[width=50pt]{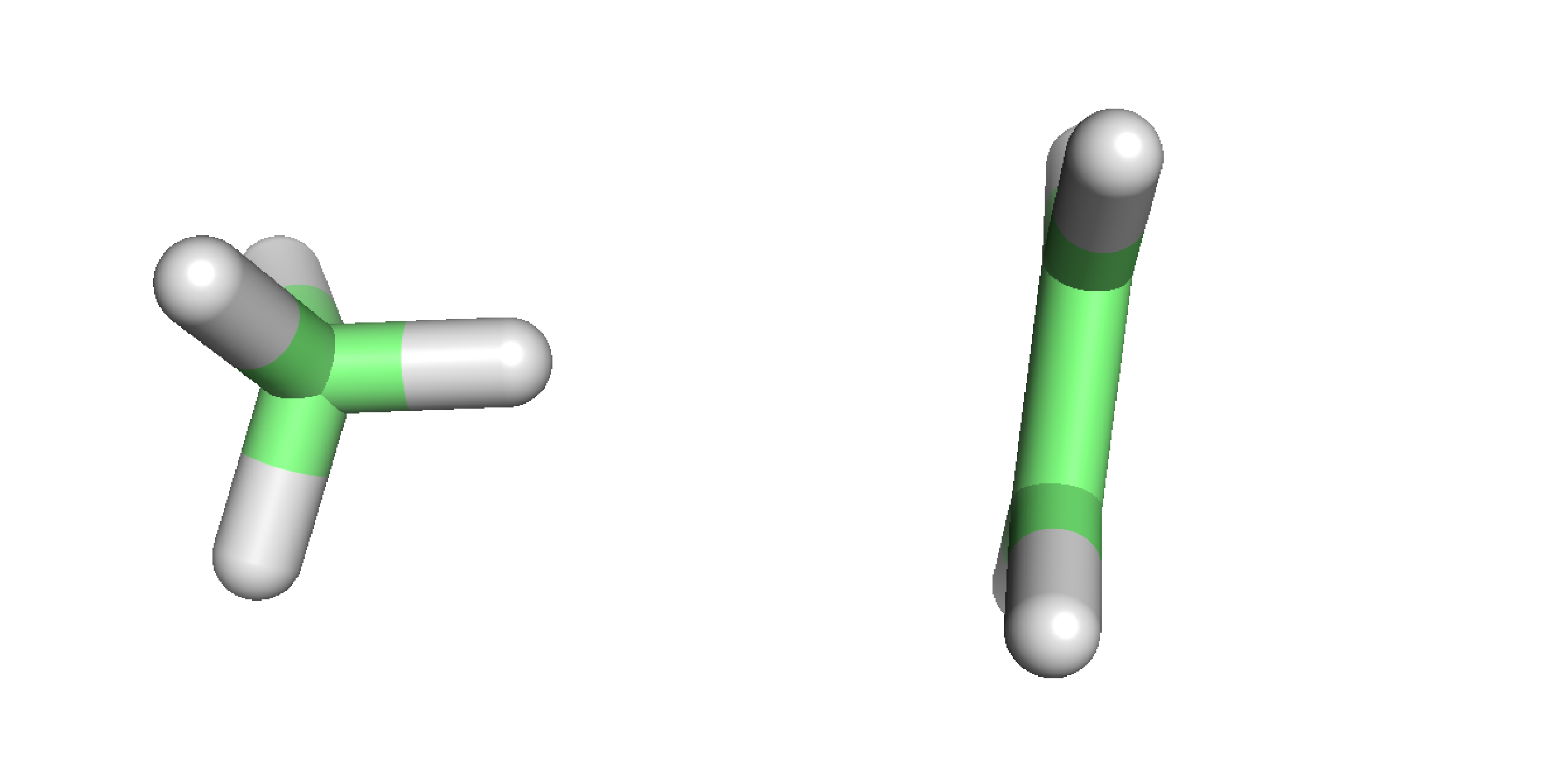}  &-1.621 & -1.70$\pm$0.08 & -0.07 & 23 & Ethene dimer & \includegraphics[width=50pt]{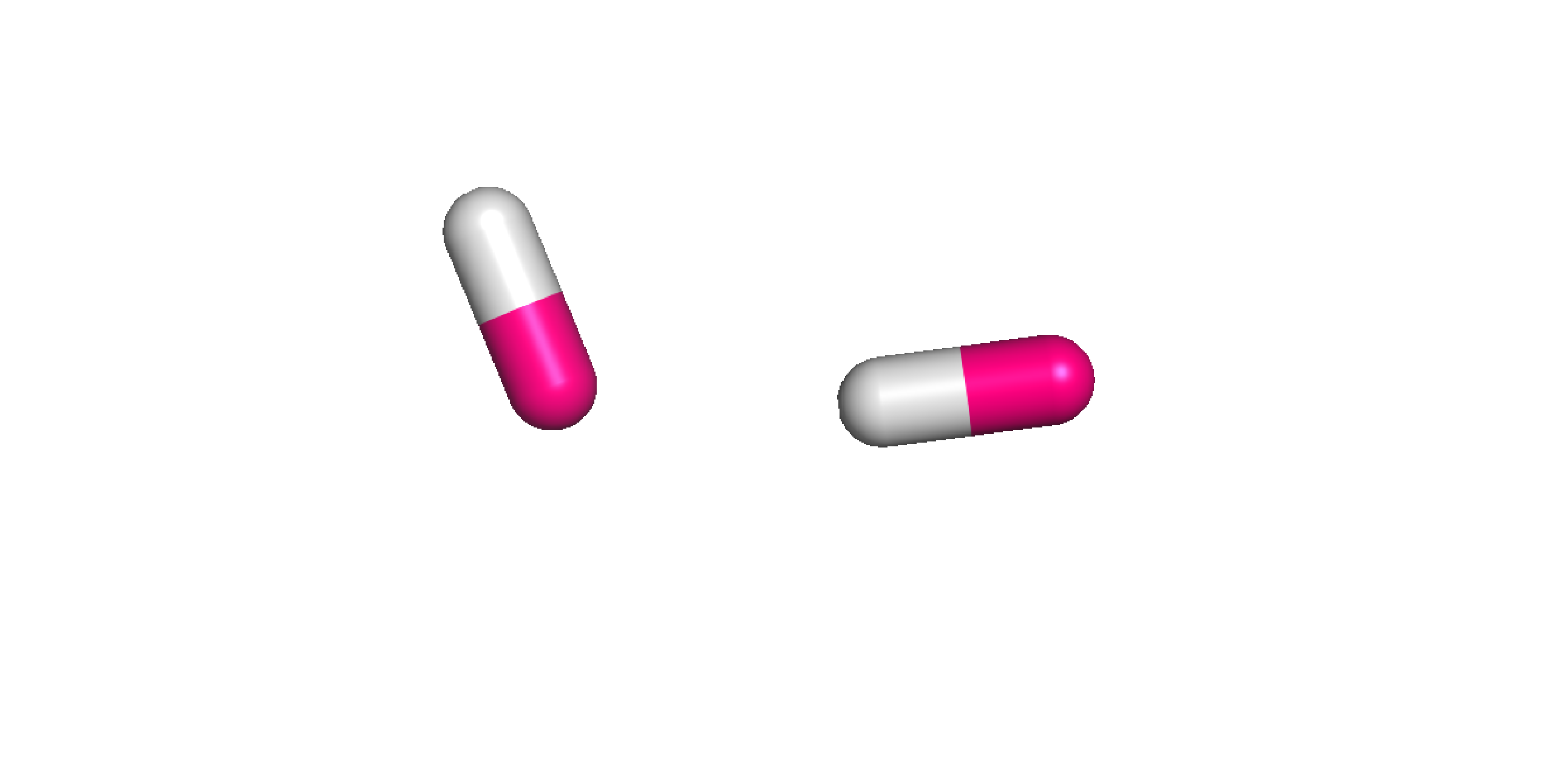}     & 0.934 & 1.06$\pm$0.03  &  0.13\\
12 & Ethyne dimer  & \includegraphics[width=50pt]{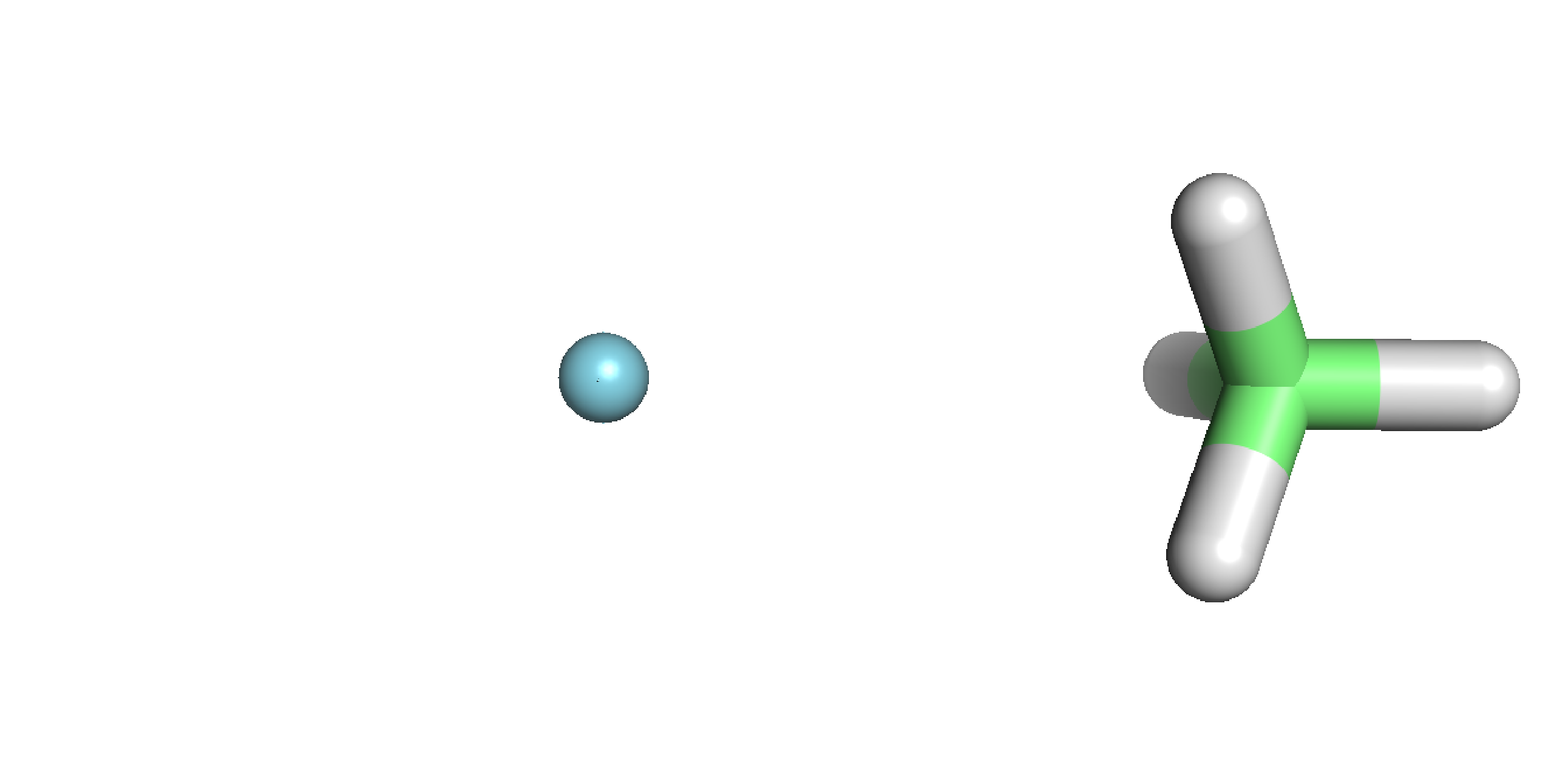}  &-1.524 & -1.74$\pm$0.07 & -0.21 & 24 & Ethyne dimer & \includegraphics[width=50pt]{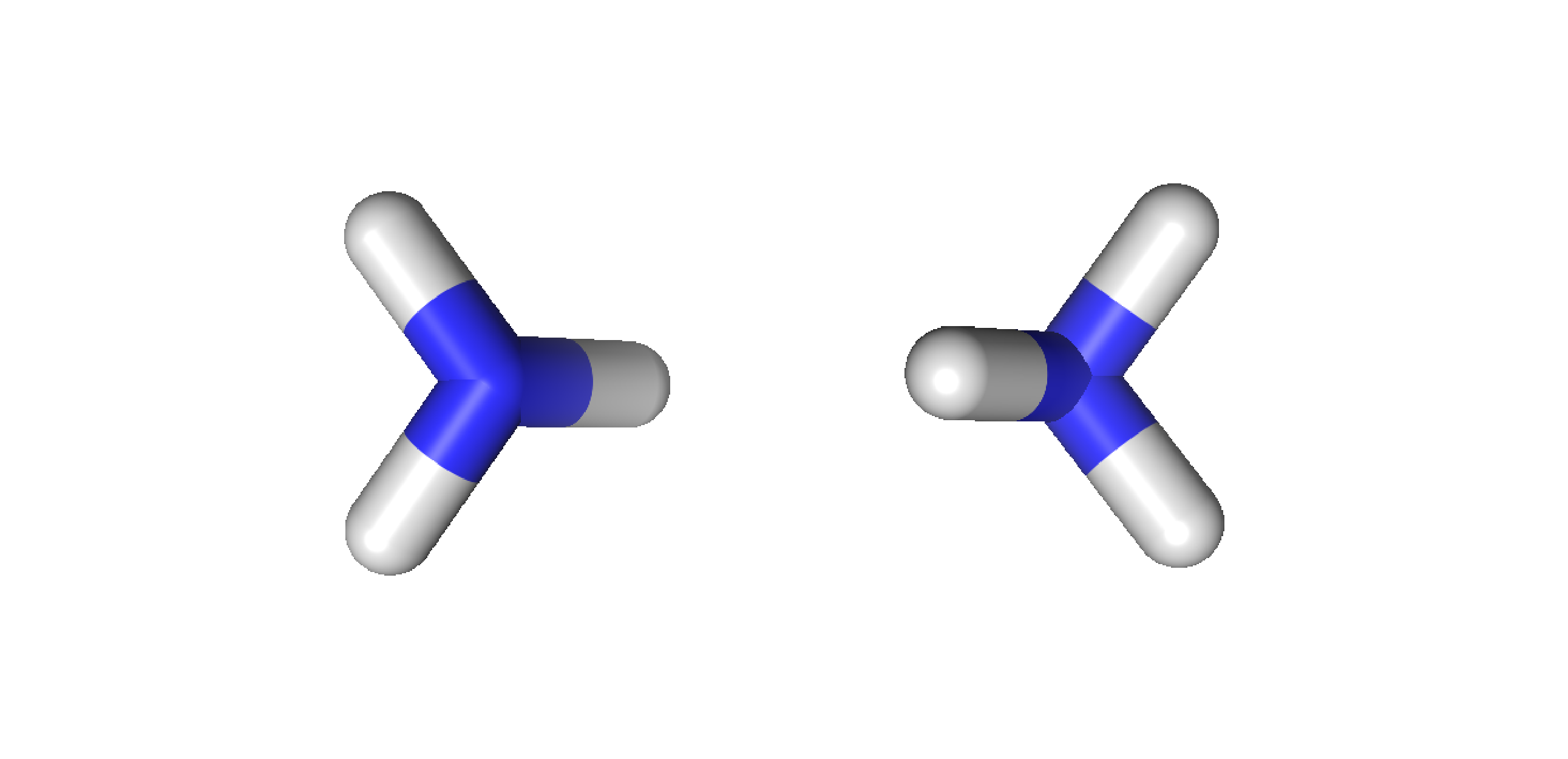}     & 1.115 & 1.23$\pm$0.03  &  0.11\\
\hline
&&&&&&&&&&ME:     &-0.05\\
&&&&&&&&&&MUE:    & 0.15\\
&&&&&&&&&&RUE:    & 12.1\\
\hline
\hline
\end{tabular}
\label{tabresA24}
\end{table*}
{The overall MUE amounts to 0.15~kcal/mol, confirming our conclusions based on Tab.~\ref{tabres}. One may observe that in general, the hydrogen-bound complexes deviate more from the reference than the mixed and dispersion-bound ones, showing the most favorable domain of the protocol applicability. The high RUE may be attributed to the high ratio between interaction energy amplitudes and noise. The overall performance of the 2tJ protocol in A24 is nevertheless very good. Some of the cases with considerable deviations were further checked using the 3tJ protocol (Tab.~\ref{tabresA24-3tJ}), which brings the interaction energies to within $\sim$0.1 kcal/mol from the reference, as expected. The HCN dimer {and the formaldehyde dimer} represent the cases where the interaction energy remains off, even with the 3tJ protocol. An analysis and fundamental understanding of these effects, beyond a scope of the current work, remains as our next goal. We tentatively assign these deviations to the presence of multiple bonds 
in combination with hydrogen bonding, possibly causing nodal surface nonlinearity effects~\cite{Rasch2013} that may lead to partial breakdown of the fixed-node error cancellation assumption~\cite{Dubecky2013}. On the other hand, the deviations still remain well below the chemical accuracy. The expected error margin of the 2tJ protocol thus remains at 0.2-0.3 kcal/mol per bond (formally).}
\begin{table}[h!]
\scriptsize
\caption{The FN-DMC interaction energies $E$ (kcal/mol) of the selected complexes from A24, obtained by the 2tJ and 3tJ protocols, compared to the CCSD(T)/CBS reference interaction energies $E_{R}$ (kcal/mol) and the corresponding differences $\Delta$.}
\newcolumntype{C}{>{\centering\arraybackslash} m{1.5cm} }
\centering
\begin{tabular}{cccccr}
\hline
\hline
Label&Complex & $E_R$ &Protocol& $E$ & $\Delta$  \\
\hline
3 & HCN dimer      & -4.745 & 2tJ & -5.09$\pm$0.08 & -0.35\\
  &                &        & 3tJ & -5.13$\pm$0.06 & -0.39\\\\
6 & Methane HF     & -1.654 & 2tJ & -1.29$\pm$0.07 & 0.37\\
  &                &        & 3tJ & -1.54$\pm$0.04 & 0.11\\\\
{9} & {Formaldehyde dimer}  &  {-4.554} &{2tJ} & {-4.90}$\pm$0.10 & {-0.35} \\
  &                &   & {3tJ} & {-4.88$\pm$0.06 } & {-0.32} \\\\
12& Ethyne dimer   & -1.524 & 2tJ & -1.74$\pm$0.07 & -0.21\\
  &                &        & 3tJ & -1.62$\pm$0.05 & -0.10\\\\
21& Ethene Ar      & -0.364 & 2tJ & -0.19$\pm$0.07 & -0.18\\
  &                &        & 3tJ & -0.22$\pm$0.04 & -0.14\\
\hline\hline
\end{tabular}
\label{tabresA24-3tJ}
\end{table}

\section{Conclusions}
The analysis of the QMC FN-DMC-based protocols and related tradeoffs provided in this work revealed a favourable  computational scheme based on a simplified explicit correlation Jastrow term that results in a reduced computational cost scaling. The tests on a number of complexes including up to a stacked DNA base pair show a good performance and nearly sub-chemical accuracy with respect to CCSD(T)/CBS, consequently enabling an easier access to reliable estimates of interaction energies in large complexes.

Since the QMC is not limited to single-reference and ``gas-phase'' complexes and the reported protocol relies on the fixed-node error cancellation, it may also find application in cases that are intrinsically difficult for mainstream-correlated wave-function approaches. These include estimates of noncovalent interactions between open-shell organometallic and/or metal-organic systems (using simple multi-determinant trial functions instead of a single Slater determinant), studies of periodic models like physisorption on metal surfaces and 2D materials, or prediction of noncovalent crystal stability~\cite{Hongo2010}.
In the domain of noncovalent interactions, the quantum Monte Carlo FN-DMC method therefore appears to be very promising for its benchmark accuracy, low-order polynomial scaling with the system complexity, low memory requirements and nearly ideal scaling across thousands of proccessors~\cite{Needs2010} in parallel supercomputing environments.

\section*{Methods}
The geometries of the studied complexes were taken from the sets S22~\cite{Jurecka2006} and A24~\cite{Rezac2013}, except for the benzene/H$_2$~\cite{Rubes2009} complex. ECPs with the corresponding basis sets developed by Burkatzki et al.~\cite{Burkatzki2007} were used throughout the work, with the exception of the H where a more recent version was used~\cite{ClaudiaPC}. The augmentation functions were taken from the corresponding Dunning bases~\cite{Dunning1989}. Single-determinant Slater-Jastrow~\cite{Jastrow1955} trial wave functions were constructed using B3LYP or HF
orbitals from {\tt GAMESS}~\cite{gamess}.
The used Schmidt-Moskowitz~\cite{Moskowitz1992} homogeneous and isotropic Jastrow factors~\cite{Jastrow1955}, including either the electron-electron and electron-nucleus terms (2tJ), or 
2tJ with electron-electron-nucleus terms in addition (3tJ), were expanded in a fixed basis set of polynomial Pad\'{e} functions~\cite{Bajdich2009rev}.
The parameters of the positive definite Jastrow factor were optimised by the Hessian driven VMC optimisation of at least 10x10 iterations
(i.e. a full VMC energy calculation after each 10 optimisation steps on a fixed walker population), using the variance optimisation~\cite{Drummond2005} or a linear combination~\cite{Umrigar2005} of energy (95\%) and variance (5\%) as a cost function. The optimised trial wave functions were subsequently used in the production FN-DMC runs
performed with a time step of 0.01/0.005~a.u. within the locality approximation~\cite{Mitas1991,Foulkes2001rev} (LA, for testing purposes) or using the \mbox{T-moves} scheme~\cite{Casula2006} for the treatment of ECPs beyond LA. The target walker populations ranged from 5k (for small systems) up to about 20k (for the largest system). 
All QMC calculations were performed using the code {\tt QWalk}~\cite{qwalk}.

The reference interaction energy for the benzene/H$_2$ complex was estimated by the basis set superposition error corrected standard CBS extrapolation technique~\cite{Jurecka2006} from \mbox{HF/aug-cc-pV5Z}, \mbox{MP2/aug-cc-pVQZ/aug-cc-pV5Z} and \mbox{CCSD(T)/aug-cc-pVQZ} results.

\footnotesize
\section*{Acknowledgements}
M.D. is grateful to Claudia Filippi for sharing an 
improved effective core potential for H and valuable discussions.
The support from the Operational Programme (OP) Research and Development (R\&D) for Innovations - European Regional Development
Fund (ERDF, project CZ.1.05/2.1.00/03.0058) and the OP Education for
Competitiveness - European Social Fund (projects CZ.1.07/2.3.00/30.0004 and CZ.1.07/2.3.00/20.0058) is gratefully acknowledged.
This work was supported by the grants P208/10/1742~(P.J.) and P208/12/G016~(M.O.), from the Czech Science Foundation.
R.D. acknowledges support from grants APVV-0207-11 and VEGA (2/0007/12).
L.M. is supported by NSF OCI-0904794, DMR-1410639 and ARO W911NF-04-D-0003-0012
grants and by XSEDE computer time allocation at TACC.
The calculations were in part performed at the Slovak infrastructure for high-performance computing (projects ITMS 26230120002 and 26210120002)
supported by the OP R\&D funded by the ERDF.
Access to computing and storage facilities owned by parties and projects contributing to the National Grid Infrastructure MetaCentrum, provided under the programme "Projects of Large Infrastructure for Research, Development, and Innovations" (LM2010005), is greatly appreciated.

\footnotesize{

\providecommand*{\mcitethebibliography}{\thebibliography}
\csname @ifundefined\endcsname{endmcitethebibliography}
{\let\endmcitethebibliography\endthebibliography}{}

}

\newpage

\begin{figure*}
\centering
\includegraphics[width=8cm]{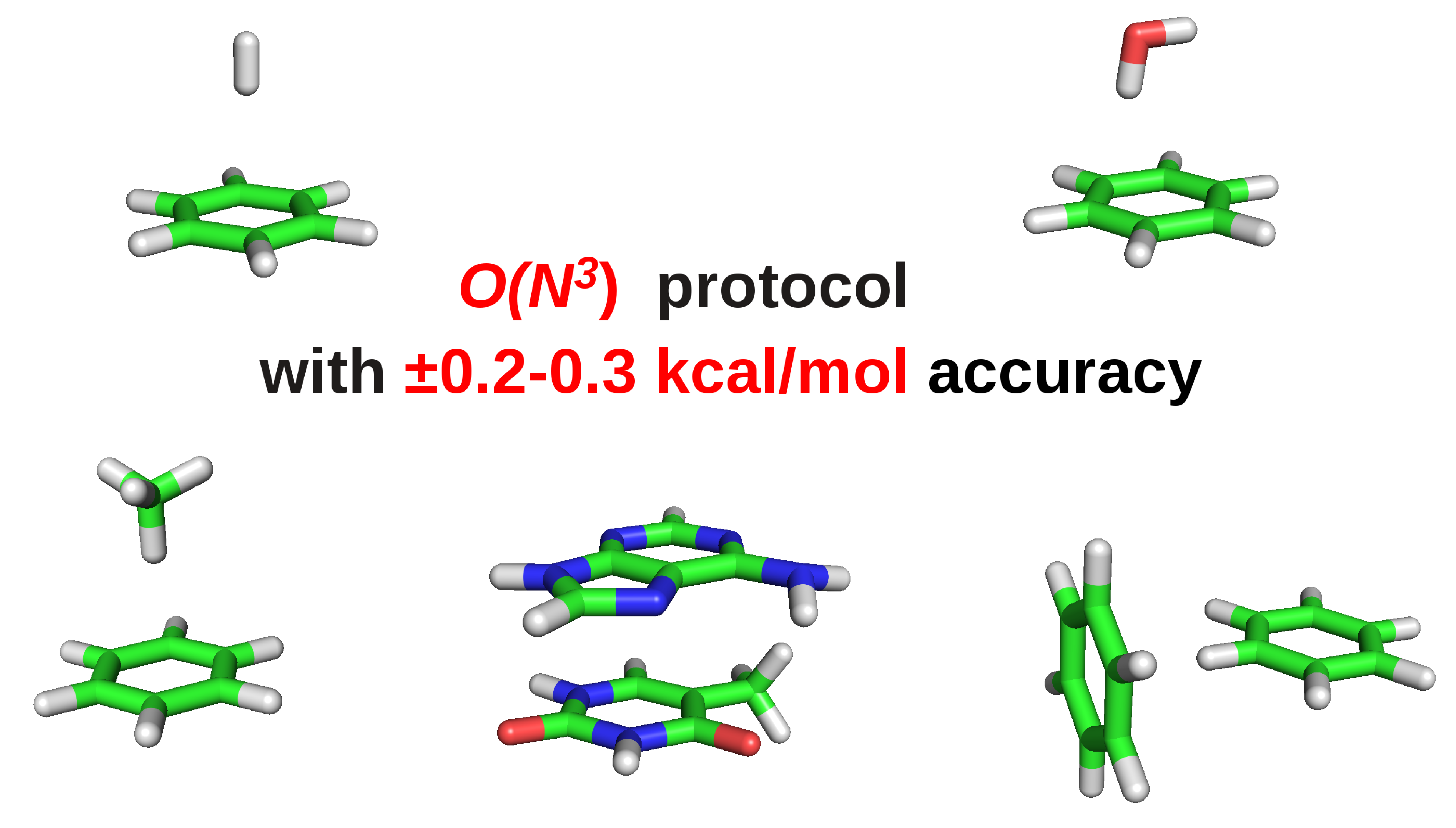}
\caption{
\textbf{Graphical TOC:}~~~A quantum Monte Carlo protocol with a favorable scaling, that attains a benchmark accuracy, is reported and extensively tested.
}

\label{fig_TOC}
\end{figure*}

\end{document}